\documentclass%[twocolumn,showpacs,preprintnumbers,amsmath,amssymb,superscriptaddress,floatfix]{revtex4}
[twocolumn,showpacs,preprintnumbers,amsmath,amssymb,floatfix]{revtex4}

\usepackage{graphicx}% Include figure files
\usepackage{dcolumn}% Align table columns on decimal point
\usepackage{bm}% bold math
\usepackage{latexsym}
\usepackage[hang,small,bf]{caption}
\usepackage[subrefformat=parens]{subcaption}
\usepackage{here}

\def\Tr{\text{Tr}} % to use \Tr instead on \matrhm{Tr}
\def\nn{\nonumber}
\newcommand{\beqn}{\begin{eqnarray}}
\newcommand{\beq}{\begin{equation}}
\newcommand{\eeqn}{\end{eqnarray}}
\newcommand{\eeq}{\end{equation}}

\newcommand{\tr}{\mathop{\rm Tr}}

\newcommand{\Kanazawa}{\affiliation{Kanazawa University, Kanazawa 920-1192, Japan}}
\newcommand{\Kochi}{\affiliation{Library and Information Technology, Kochi University, Kochi 780-8520, Japan}}
\newcommand{\IHEP}{\affiliation{NRC "Kurchatov Institute" -IHEP, 142281, Protvino, Russia \\
School of Biomedicine, Far Eastern Federal University, Vladivostok 690950, Russia }}

\begin{document}

\title{A new scheme for color confinement and violation of the non-Abelian Bianchi identities}
\author{Tsuneo Suzuki}
\email[e-mail:]{suzuki04@staff.kanazawa-u.ac.jp}
\Kanazawa
\author{Katsuya Ishiguro}
\Kochi
\author{Vitaly Bornyakov}
\IHEP

\date{\today}% It is always \today, today,

\begin{abstract}
% revised
%%A new scheme for color confinement in QCD due to violation of the non-Abelian Bianchi identities is proposed.
A new scheme for color confinement in QCD due to violation of the non-Abelian Bianchi identities proposed earlier is revised. 
%%%%%
The violation of the non-Abelian Bianchi identities (VNABI) $J_{\mu}$  is equal to Abelian-like monopole currents $k_{\mu}$ defined by the violation of the Abelian-like Bianchi identities.   Although VNABI is an adjoint operator satisfying the covariant conservation law $D_{\mu}J_{\mu}=0$, it satisfies,
at the same time, the Abelian-like conservation law $\partial_{\mu}J_{\mu}=0$.  There are $N^2-1$ conserved magnetic charges in  $SU(N)$ QCD. The charge of each component of VNABI  is assumed to satisfy the Dirac quantization condition. 
%%%%%
Each color component of the non-Abelian electric field $E^a$ is squeezed by the corresponding color component of the solenoidal current $J^a_{\mu}$. Then only the color singlets alone can survive as a physical state and non-Abelian color confinement is realized.  This confinement picture is completely new in comparison with the previously studied monopole confinement scenario based on an Abelian projection after some partial gauge-fixing, where Abelian neutral states can survive as physical.  

To check if the scenario is realized in nature, numerical studies are done in the framework of lattice field theory
 by adopting pure $SU(2)$ gauge theory for simplicity. Considering $J_{\mu}(x)=k_{\mu}(x)$ in the continuum formulation, we adopt
an Abelian-like definition of a monopole following DeGrand-Toussaint  as a lattice version of VNABI,
since the Dirac quantization condition of the magnetic charge is satisfied on lattice partially.
 To reduce severe lattice
artifacts, we introduce various techniques of smoothing the thermalized vacuum.  Smooth gauge fixings such as
the maximal center gauge (MCG), block-spin transformations of Abelian-like monopoles and extraction of physically
important infrared long monopole loops are adopted.  We also employ the tree-level tadpole improved gauge action of $SU(2)$ gluodynamics.
With these various improvements, we measure the density of lattice VNABI: 
$\rho(a(\beta),n)=\sum_{\mu,s_n}\sqrt{\sum_a(k_{\mu}^a(s_n))^2}/(4\sqrt{3}V_nb^3)$,
where $k_{\mu}^a(s_n)$ is an $n$ blocked monopole in the color direction $a$, $n$ is the number of blocking steps, $V_n=V/n^4$ ($b=na(\beta)$) is the lattice  volume (spacing) of the blocked lattice. Beautiful and convincing scaling behaviors are seen
when we plot the density $\rho(a(\beta),n)$ versus $b=na(\beta)$. A single universal curve $\rho(b)$ is found from $n=1$ to $n=12$,
which suggests that $\rho(a(\beta),n)$ is a function of $b=na(\beta)$ alone.  The universal curve seems independent of a gauge fixing procedure used to smooth the lattice vacuum since the scaling is obtained in all gauges adopted.  The scaling, if it exists also for $n\to\infty$,
shows that the lattice definition of VNABI has the continuum limit and the new confinement scenario is realized.

\end{abstract}

\maketitle

\section{Introduction}
Color confinement in  quantum chromodynamics (QCD) 
is still an important unsolved  problem~\cite{CMI:2000mp}. 

As a picture of color confinement, 't~Hooft~\cite{tHooft:1975pu} and Mandelstam~\cite{Mandelstam:1974pi} conjectured that the QCD vacuum is a kind of a magnetic superconducting state caused by condensation of magnetic monopoles and  an effect dual to the Meissner effect works to confine color charges. 
However, in contrast to SUSY QCD
~\cite{Seiberg:1994rs} or Georgi-Glashow 
model~\cite{'tHooft:1974qc,Polyakov:1976fu} with scalar fields,
to find color magnetic monopoles which
condense is not straightforward in QCD. 

An interesting idea to realize this conjecture is to project QCD to the Abelian 
maximal torus group by a partial (but singular) gauge fixing~\cite{tHooft:1981ht}. 
In $SU(3)$ QCD, the maximal torus group is  Abelian $U(1)^2$. 
Then color magnetic monopoles appear as a topological object.
Condensation of the monopoles  causes  the dual Meissner 
effect~\cite{Ezawa:1982bf,Suzuki:1988yq,Maedan:1988yi}.

\par
Numerically, an Abelian projection in non-local 
gauges such as the maximally Abelian (MA)
gauge~\cite{Suzuki:1983cg,Kronfeld:1987ri,Kronfeld:1987vd} 
has been found to support the Abelian confinement scenario
beautifully~\cite{Suzuki:1992rw,Singh:1993jj,Chernodub:1997ay,Bali:1997cp,
Suzuki:1998hc,Koma:2003gq,Koma:2003hv}. Also the  Abelian dominance and the dual Meissner effect are observed  clearly in local unitary gauges such as $F12$ and Polyakov (PL) gauges~\cite{Sekido:2007mp}.  

\par
However, although numerically interesting, the idea of Abelian projection\cite{tHooft:1981ht} is theoretically very unsatisfactory. 1) In non-perturabative QCD, 
any gauge-fixing is  not necessary at all. There are infinite ways of such a partial gauge-fixing and whether the 't Hooft scheme is gauge independent or 
not is not known. 2)  After an Abelian projection, only one (in $SU(2)$) or two (in $SU(3)$) gluons are photon-like with respect to the residual $U(1)$ or $U(1)^2$ symmetry and the other gluons are massive charged matter fields. Such an asymmetry among gluons is  unnatural. 3) How to construct Abelian monopole operators in a gauge-independent way in terms of original gluon fields is not clear at all.

\vspace{.5cm}

In this paper, we propose a new theoretical scheme for color confinement based on the dual Meissner effect which is free from the above problems.   The idea was first expressed by one of the authors (T.S.) in Ref.\cite{Suzuki:2014wya} and was extended in Ref.\cite{SIB201711}. However, 
the proofs of the Dirac quantization condition of $g_m^a$ in $SU(2)$ and $SU(3)$ shown in Refs.\cite{Suzuki:2014wya,SIB201711} are incorrect.   Without knowing the explicit form of the gauge-field configuration corresponding to VNABI, it is impossible to prove the Dirac quantization condition theoretically. Since the authors expect that VNABI play an important role in color confinement, the Dirac quantization conditions for $g_m^a$ in $SU(2)$ and $SU(3)$ are assumed. Also   the simultaneous diagonalization of VNABI $J_{\mu}$ for all $\mu$ can
not be proved from the Coleman-Mandula theorem\cite{Coleman} and Lorentz invariance contrary to the assertion in Ref.\cite{SIB201711}.  
When the simultaneous diagonalization of $J_{\mu}$ for all $\mu$ is assumed, the condensation of $J_\mu$ and electric color invariance of the confinement vacuum can be compatible.

Then to check if the above scheme is realized in nature, we study the proposal in the framework of the non-Abelian lattice gauge theory. For simplicity we adopt pure $SU(2)$ lattice gauge theory.
First considering $J_{\mu}(x)=k_{\mu}(x)$ in the continuum, we define VNABI on lattice as an Abelian-like monopole following DeGrand-Toussaint\cite{DeGrand:1980eq}. Then as a most important point to be clarified, we are going to study if the lattice VNABI has the non-trivial continuum limit,
namely if the scaling of the density exists. 

The lattice monopoles exist as a closed loop due to the current conservation law.
As shown later explicitly, monopole closed loops are 
contaminated by lattice artifacts. Hence it is absolutely necessary to
introduce various techniques avoiding such large lattice artifacts in order to
analyse especially such a quantity as the monopole density, since all lattice artifacts contribute positively to the density.
We introduce various techniques of smoothing the thermalized vacuum.  Smooth gauge fixings such as  the maximal center gauge (MCG)\cite{DelDebbio:1996mh,DelDebbio:1998uu},
block-spin transformations of Abelian-like monopoles and extraction of physically important infrared long monopoles are taken into account.
We also employ the tree-level tadpole improved gauge action.

\section{A new confinement scheme based on VNABI\label{Sec2}}
\subsection{Equivalence of  $J_{\mu}$ and $k_{\mu}$}
First of all, we prove that the Jacobi identities of covariant derivatives lead us to conclusion that  violation of the non-Abelian Bianchi identities (VNABI) $J_{\mu}$ is nothing but an Abelian-like monopole $k_{\mu}$ defined by violation of the Abelian-like Bianchi identities without gauge-fixing. 
 Define a covariant derivative operator $D_{\mu}=\partial_{\mu}-igA_{\mu}$. The Jacobi identities are expressed as 
\begin{eqnarray}
\epsilon_{\mu\nu\rho\sigma}[D_{\nu},[D_{\rho},D_{\sigma}]]=0. \label{eq-Jacobi}
\end{eqnarray}
By direct calculations, one gets
\begin{eqnarray*}
[D_{\rho},D_{\sigma}]&=&[\partial_{\rho}-igA_{\rho},\partial_{\sigma}-igA_{\sigma}]\\
&=&-ig(\partial_{\rho}A_{\sigma}-\partial_{\sigma}A_{\rho}-ig[A_{\rho},A_{\sigma}])+[\partial_{\rho},\partial_{\sigma}]\\
&=&-igG_{\rho\sigma}+[\partial_{\rho},\partial_{\sigma}],
\end{eqnarray*}
where the second commutator term of the partial derivative operators can not be discarded, since gauge fields may contain a line singularity. Actually, it is the origin of the violation of the non-Abelian Bianchi identities (VNABI) as shown in the following. The non-Abelian Bianchi identities and the Abelian-like Bianchi identities are, respectively: $D_{\nu}G^{*}_{\mu\nu}=0$ and $\partial_{\nu}f^{*}_{\mu\nu}=0$.
The relation $[D_{\nu},G_{\rho\sigma}]=D_{\nu}G_{\rho\sigma}$ and the Jacobi identities (\ref{eq-Jacobi}) lead us to
\begin{eqnarray}
D_{\nu}G^{*}_{\mu\nu}&=&\frac{1}{2}\epsilon_{\mu\nu\rho\sigma}D_{\nu}G_{\rho\sigma} \nn\\
&=&-\frac{i}{2g}\epsilon_{\mu\nu\rho\sigma}[D_{\nu},[\partial_{\rho},\partial_{\sigma}]]\nn\\
&=&\frac{1}{2}\epsilon_{\mu\nu\rho\sigma}[\partial_{\rho},\partial_{\sigma}]A_{\nu}\nn\\
&=&\partial_{\nu}f^{*}_{\mu\nu}, \label{eq-JK}
\end{eqnarray}
where $f_{\mu\nu}$ is defined as $f_{\mu\nu}=\partial_{\mu}A_{\nu}-\partial_{\nu}A_{\mu}=(\partial_{\mu}A^a_{\nu}-\partial_{\nu}A^a_{\mu})\sigma^a/2$. Namely Eq.(\ref{eq-JK}) shows that the violation of the non-Abelian Bianchi identities is equivalent to that of the Abelian-like Bianchi identities.

Denote the violation of the non-Abelian Bianchi identities as  $J_{\mu}$:
\begin{eqnarray}
J_{\mu} &=& \frac{1}{2}J_{\mu}^a\sigma^a
=D_{\nu}G^*_{\mu \nu}. \label{nabi}
\end{eqnarray}
Eq.(\ref{nabi}) is gauge covariant and therefore a non-zero  $J_{\mu}$ is a gauge-invariant property. An Abelian-like monopole $k_{\mu}$ without any gauge-fixing is defined as the violation of the Abelian-like Bianchi identities:
\begin{eqnarray}
k_{\mu}=\frac{1}{2}k_{\mu}^a\sigma^a&=& \partial_{\nu}f^*_{\mu\nu}
=\frac{1}{2}\epsilon_{\mu\nu\rho\sigma}\partial_{\nu}f_{\rho\sigma}. \label{ab-mon}
\end{eqnarray}
Eq.(\ref{eq-JK}) shows that
\begin{eqnarray}
J_{\mu}=k_{\mu}. \label{JK}
\end{eqnarray}
 
Several comments are in order.
\begin{enumerate}
\item Eq.(\ref{JK}) can be considered as 
a special case 
of the important relation derived by Bonati et al.\cite{Bonati:2010tz} in the framework of an Abelian projection to a simple case without any Abelian projection.
Actually it is possible to prove directly without the help of the Jacobi identities
\begin{eqnarray*}
J_{\mu}^a-k_{\mu}^a&=& \Tr\sigma^a D_{\nu}G^{*}_{\mu\nu}-\partial_{\nu}f^{*a}_{\mu\nu} \\
&=&-ig\Tr\sigma^a[A_{\nu}, G^{*}_{\mu\nu}]\\
&&-ig\epsilon_{\mu\nu\rho\sigma}\Tr\sigma^a[\partial_{\nu}A_{\rho}, A_{\sigma}]\\
&=&0. 
\end{eqnarray*}
\item
VNABI $J_{\mu}$ transforms as an adjoint operator, so that does the Abelian-like monopole current $k_{\mu}$. This can be proved also directly. Consider a regular gauge transformation
\begin{eqnarray*}
A'_{\mu}&=&VA_{\mu}V^{\dag}-\frac{i}{g}\partial_{\mu}VV^{\dag}.
\end{eqnarray*}
Then
\begin{eqnarray}
k'_{\mu}&=&\epsilon_{\mu\nu\rho\sigma}\partial_{\nu}\partial_{\rho}A'_{\sigma}\nn\\
&=&\epsilon_{\mu\nu\rho\sigma}\partial_{\nu}\partial_{\rho}(VA_{\sigma}V^{\dag}-\frac{i}{g}\partial_{\sigma}VV^{\dag})\nn\\
&=&V(\epsilon_{\mu\nu\rho\sigma}\partial_{\nu}\partial_{\rho}A_{\sigma})V^{\dag}\nn\\
&=&Vk_{\mu}V^{\dag}.\label{vkv}
\end{eqnarray}

\item
The above equivalence shows VNABI is essentially Abelian-like. It was already argued that singularities of gauge fields corresponding to VNABI must be Abelian\cite{DiGiacomo:2008wh}, although the reasoning is different.
\item The covariant conservation law $D_{\mu}J_{\mu}=0$ is proved as follows\cite{Bonati:2010tz}:
\begin{eqnarray}
D_{\mu}J_{\mu}&=&D_{\mu}D_{\nu}G^*_{\nu\mu}
=\frac{ig}{2}[G_{\nu\mu},G^*_{\nu\mu}]\nn\\
&=&\frac{ig}{4}\epsilon_{\nu\mu\rho\sigma}[G_{\nu\mu},G_{\rho\sigma}]
=0,  \label{NA-cons}
\end{eqnarray}
where 
\begin{eqnarray}
\partial_{\mu}\partial_{\nu}G^{*}_{\mu\nu}=0 \label{PNA}
\end{eqnarray}
 is used.
The Abelian-like monopole  satisfies the Abelian-like  conservation law
\begin{eqnarray}
\partial_{\mu}k_{\mu}=\partial_{\mu}\partial_{\nu}f^{*}_{\mu\nu}=0\label{PAA}
\end{eqnarray}
due to the antisymmetric property of the Abelian-like field strength\cite{Arafune:1974uy}. Hence VNABI satisfies also the same Abelian-like conservation law 
\begin{eqnarray}
\partial_{\mu}J_{\mu}=0. \label{A-cons}
\end{eqnarray}
Both Eqs.(\ref{NA-cons}) and (\ref{A-cons}) are compatible, since
the difference between both quantities
\begin{eqnarray}
[A_{\mu}, J_{\mu}]&=&\frac{1}{2}\epsilon_{\mu\nu\rho\sigma}[A_{\mu},\partial_{\nu}f_{\rho\sigma}]\nn\\
&=&\epsilon_{\mu\nu\rho\sigma}[A_{\mu}, \partial_{\nu}\partial_{\rho}A_{\sigma}]\nn\\
&=&-\frac{1}{2}\epsilon_{\mu\nu\rho\sigma}\partial_{\nu}\partial_{\mu}[A_{\rho},A_{\sigma}]\nn\\
&=&\frac{i}{g}(\partial_{\mu}\partial_{\nu}G^*_{\mu\nu}-\partial_{\mu}\partial_{\nu}f^*_{\mu\nu})\nn\\
&=& 0\nn,
\end{eqnarray}
where (\ref{PNA}) and (\ref{PAA}) are used.
 Hence the Abelian-like conservation relation (\ref{A-cons}) is also gauge-covariant.

\item The Abelian-like conservation relation (\ref{A-cons}) gives us three conserved magnetic charges in the case of color $SU(2)$ and $N^2-1$ charges in the case of color $SU(N)$. But these are kinematical  relations coming from the derivative with respect to the divergence of an antisymmetric tensor~\cite{Arafune:1974uy}.  The number of conserved charges is different from that of the Abelian projection scenario~\cite{tHooft:1981ht}, where only $N-1$ conserved charges exist in the case of color $SU(N)$.
\end{enumerate}

\begin{table*}
\caption{\label{comp}Comparison between the 'tHooft Abelian projection studies and the present work in $SU(2)$ QCD.  $\hat{\phi}'=V_p^{\dag}\sigma_3 V_p$, where $V_p$ is a partial gauge-fixing matrix of an Abelian projection. $(u_c, d_c)$ is a color-doublet quark pair. MA means maximally Abelian. }
\begin{ruledtabular}
\begin{tabular}{|c|c|c|c|}
  &\multicolumn{2}{c|}{The 'tHooft Abelian projection scheme}  &  This work and Refs.\cite{Suzuki:2007jp,Suzuki:2009xy} \ \   \\
  &Previous works\cite{Suzuki:1983cg,Kronfeld:1987ri,Kronfeld:1987vd,Suzuki:1992rw,Singh:1993jj,Chernodub:1997ay,Bali:1997cp,Suzuki:1998hc,Koma:2003gq,Koma:2003hv,Sekido:2007mp} & Reference \cite{Bonati:2010tz} & \\
  \hline
  Origin of  $k_{\mu}$ & A singular gauge transformation& $k_{\mu}=\Tr J_{\mu}\hat{\phi}'$ & $k_{\mu}^a=J_{\mu}^a$\\
  \hline
No. of conserved  $k_{\mu}$   & \multicolumn{2}{c|}{$1$}   &  $3$  \\ 
Role of $A^a_{\mu}$ &\multicolumn{2}{c|}{One photon $A^3_{\mu}$ with $k^3_{\mu}$ $+$ 2 massive $A^{\pm}_{\mu}$} &  Three gluons $A_{\mu}^a$ with $k_{\mu}^a$ \\
Flux squeezing&\multicolumn{2}{c|}{One electric field $E_{\mu}$ }& Three electric fields $E^a_{\mu}$\\
   \hline
 Number of physical mesons &\multicolumn{2}{c|}{ 2 Abelian neutrals, $\bar{u}_cu_c$ and $\bar{d}_cd_c$}& 1 color singlet $\bar{u}_cu_c+\bar{d}_cd_c$\\
   \hline
 Expected confining vacuum &\multicolumn{2}{c|}{Condensation of Abelian monopoles }& Condensation of color-invariant $\lambda_{\mu}$\cite{Suzuki:1988yq}\\
 \hline 
Privileged gauge choice   &  A singular gauge &  MA gauge  & No need of gauge-fixing   \\
\end{tabular}
\end{ruledtabular} 
\end{table*}

\vspace{.5cm}
\subsection{Proposal of the vacuum in the confinement phase}
Now we propose a new mechanism of color confinement in which VNABI $J_{\mu}$ play an important role in the vacuum. 
For the scenario to be realized, we make two assumptions concerning the property of VNABI. 
\begin{enumerate}
  \item If VNABI are important physically, they must satify the Dirac quantization condition between the gauge coupling $g$ and the magnetic charge $g_m^a$ for $a=1,2,3$ in $SU(2)$ and $a=1\sim 8$ in $SU(3)$. 
Since we do not know theoretically the property of VNABI, we have to assume the Dirac qunatization conditions:
\begin{eqnarray*}
gg_m^a=4\pi n^a,
\end{eqnarray*}
where $n^a$ is an integer. 
\item The vacuum in the color confinement phase should be electric color invariant. Since VNABI transform as an adjoint operator, we have to extract electric color invariant but magnetically charged quantity from VNABI. One possible way it to assume that VNABI satisfy
\begin{eqnarray*}
[J_\mu (x),J_{\nu\neq\mu}]=0
\end{eqnarray*}
which make it possible to diagonalize VNABI $J_\mu$ simultaneously 
for all $\mu$. 
At present, the authors do not know if the second assumption is the only way to have the magnetically charged but electrically neutral vacuum in the confinement phase.
\end{enumerate}

Using the above assumption, VNABI can be diagonalized by a unitary matrix $V_d(x)$ as follows:
\begin{eqnarray*}
V_d(x)J_{\mu}(x)V_d^{\dag}(x)=\lambda_{\mu}(x)\frac{\sigma_3}{2},
\end{eqnarray*}
where $\lambda_{\mu}(x)$ is the eigenvalue of $J_{\mu}(x)$ and is then color invariant but magnetically charged. 
% deleted
%Note that $V_d(x)$ does not depend on $\mu$ due to the Coleman-Mandula theorem\cite{Coleman}
%\cite{footnote1}.
%\footnote{Applying  the Coleman-Mandula theorem  to QCD, I note that Kugo-Ojima\cite{Kugo} showed a manifestly covariant and local canonical operator formalism of non-Abelian gauge theories. Although introducing an indefinite metric is inevitable, the unitarity of the physical $S$-matrix is proved.  Moreover the string-like behavior existing in  gauge fields in the case of color confinement is not rejected in the framework.}. 
Then one gets
\begin{eqnarray}
\Phi(x)&\equiv& V_d^{\dag}(x)\sigma_3 V_d(x) \label{Phi}\\
J_{\mu}(x)&=&\frac{1}{2}\lambda_{\mu}(x)\Phi(x), \label{lambda}\\
\sum_a (J_{\mu}^a(x))^2&=&\sum_a (k_{\mu}^a(x))^2=(\lambda_{\mu}(x))^2. \label{Eigen}
\end{eqnarray}
Namely the color electrically charged part and the magnetically charged part are separated out. From (\ref{lambda}) and (\ref{A-cons}), one gets
\begin{eqnarray}
\partial_{\mu}J_{\mu}(x)&=&\frac{1}{2}(\partial_{\mu}\lambda_{\mu}(x)\Phi(x) + \lambda_{\mu}(x)\partial_{\mu}\Phi(x))\nn\\
&=& 0.
\end{eqnarray}
Since $\Phi(x)^2=1$, 
\begin{eqnarray*}
\partial_{\mu}\lambda_{\mu}(x)&=&-\frac{1}{2}\lambda_{\mu}(x)(\Phi(x)\partial_{\mu}\Phi(x)+\partial_{\mu}\Phi(x)\Phi(x))\\
&=&0.
\end{eqnarray*}
Hence the eigenvalue $\lambda_{\mu}$ itself satisfies the Abelian conservation rule. 

Furthermore, when use is made of (\ref{vkv}), it is possible to prove that 
\begin{eqnarray}
\frac{1}{2}\epsilon_{\mu\nu\rho\sigma}\partial_{\nu}f'_{\mu\nu}(x)
&=&\lambda_{\mu}(x)\frac{\sigma_3}{2},\label{lambda3}
\end{eqnarray}
where 
\begin{eqnarray*}
f'_{\mu\nu}(x)&=&\partial_{\mu}A'_{\nu}(x)-\partial_{\nu}A'_{\mu}(x)\\
A'_{\mu}&=&V_dA_{\mu}V^{\dag}_d-\frac{i}{g}\partial_{\mu}V_dV_d^{\dag},\\
        &\equiv& \frac{A^{'a}_{\mu}\sigma^a}{2}.
\end{eqnarray*}
Namely,
\begin{eqnarray}
\frac{1}{2}\epsilon_{\mu\nu\rho\sigma}\partial_{\nu}f^{'1,2}_{\rho\sigma}(x)(x)&=&0\\
\frac{1}{2}\epsilon_{\mu\nu\rho\sigma}\partial_{\nu}f^{'3}_{\rho\sigma}(x)(x)&=&\lambda_{\mu}(x).\label{eq:Lambda}
\end{eqnarray}
The singularity appears only in the diagonal component of the gauge field $A'_{\mu}$. 

% the following is deleted
%If one considers for large $r$
% \begin{eqnarray*}
%A'_{\mu}&\to&\Omega\partial_\mu\Omega^{\dag}/ig,\\ 
%\hat{\phi}&=&\hat{\phi}^i\sigma^i
%=\Omega\sigma^3\Omega^{\dag},
%\end{eqnarray*}
%one can easily see from (\ref{eq:Lambda}) and (\ref{GM}) that the magnetic charge from the eigenvalue $\lambda_{\mu}$ also satisfies the Dirac quantization condition  (\ref{g-gm}).

It is very interesting to see that $f^{'3}_{\mu\nu}(x)$ is actually the gauge invariant 'tHooft tensor\cite{'tHooft:1974qc}:
\begin{eqnarray*}
f^{'3}_{\mu\nu}(x)=\tr\Phi(x)G_{\mu\nu}(x)+\frac{i}{2g}\tr\Phi(x)D_{\mu}\Phi(x)D_{\nu}\Phi(x),
\end{eqnarray*}
in which the field $\Phi(x)$ (\ref{Phi}) plays a role of the scalar Higgs field in Ref.\cite{'tHooft:1974qc}. To be noted is that the field $\Phi(x)$ (\ref{Phi}) is determined uniquely by VNABI itself in the gluodynamics without any Higgs field.
In this sense, our scheme can be regarded as a special Abelian projection scenario with the partial gauge-fixing condition where $J_{\mu}(x)$ are diagonalized. 
The condensation of the gauge-invariant magnetic currents $\lambda_{\mu}$  does not give rise to a spontaneous breaking of the color electric symmetry. 
Condensation of the color invariant  magnetic currents $\lambda_{\mu}$ may be a key mechanism of  the physical confining vacuum\cite{Suzuki:1988yq, Maedan:1988yi}. 

The main difference between our new scheme and previous Abelian projection schemes is that in the former there exist  $N^2-1$ conserved magnetic currents squeezing $N^2-1$ color electric fields and color ( not charge) confinement is shown explicitly, whereas in the latter, there exists only $N-1$ conserved currents giving charge confinement. In our scheme, the $N^2-1$ conserved magnetic currents are degenerate in the vacuum to $N-1$ color-invariant currents corresponding to the eigenvalues.
To show the difference of this scheme from the previous 'tHooft Abelian projection with some partial gauge-fixing, we show  Table~\ref{comp} in which typical different points are written.

Let us make a comment  here on the relation  derived by Bonati et al.\cite{Bonati:2010tz}:
\begin{eqnarray}
k^{AB}_{\mu}(x)&=&\Tr\{J_{\mu}(x)\Phi^{AB}(x)\}, \label{kab}
\end{eqnarray}
where $k^{AB}_{\mu}(x)$ is an Abelian monopole, $\Phi^{AB}(x)=V^{\dag}_{AB}(x)\sigma_3V_{AB}(x)$ and $V_{AB}(x)$ is a partial gauge-fixing matrix in some Abelian projection  like the MA gauge. Making use of Eq.(\ref{lambda}), we get
\begin{eqnarray}
k^{AB}_{\mu}(x)&=&\lambda_{\mu}(x)\tilde{\Phi}^3(x), 
\end{eqnarray}
where 
\begin{eqnarray*}
\tilde{\Phi}(x)&=&V_{AB}(x)V_d^{\dag}(x)\sigma_3V^{\dag}_{AB}(x)V_d(x)\\
&=&\tilde{\Phi}^a(x)\sigma^a.
\end{eqnarray*}
The relation (\ref{kab}) is important, since existence of an Abelian monopole in any Abelian projection scheme is guaranteed by that of VNABI $J_{\mu}$ in the continuum limit.
Hence if in any special gauge such as MA gauge, Abelian monopoles remain non-vanishing in the continuum as suggested by many numerical data~\cite{Suzuki:1992rw,Singh:1993jj,Chernodub:1997ay,Bali:1997cp,
Suzuki:1998hc,Koma:2003gq,Koma:2003hv}, VNABI also remain non-vanishing in the continuum.

\vspace{.5cm}

\section{Lattice numerical study of the continuum limit}
\subsection{Definition of VNABI on lattice}
Let us try to define VNABI on lattice.  
In the previous section, VNABI $J_{\mu}(x)$
 is shown to be equivalent in the continuum limit to
the violation of the Abelian-like Bianchi identities
$J_{\mu}(x)=k_{\mu}(x)$.

On lattice, we have to define a quantity which leads us to the above VNABI in the continuum limit. There are two possible definitions which lead us to the above VNABI in the naive continuum limit. One  is a quantity keeping the adjoint transformation property under the lattice $SU(2)$ gauge transformation $V(s)$:
\begin{eqnarray*}
U(s,\mu)^{'}=V(s)U(s,\mu)V^{\dag}(s+\mu).
\end{eqnarray*}
Here $U(s, \mu)$ is a lattice gauge link field.
Such a quantity was proposed in Ref\cite{Skala:1996ar}:
\begin{eqnarray*}
J_{\mu}(s)&\equiv&\frac{1}{2}\big(U(s,\nu)U_{\mu\nu}(s+\nu)U^{\dag}(s,\nu)-U_{\mu\nu}(s)\big),\\
U_{\mu\nu}(s)&\equiv&U(s,\mu)U(s+\mu,\nu)U^{\dag}(s+\nu,\mu)U^{\dag}(s,\nu)
\end{eqnarray*}
where $U_{\mu\nu}(s)$ is a plaquette variable corresponding to the non-Abelian field strength. This transforms as an adjoint operator:
\begin{eqnarray}
J_{\mu}^{'}(s)=V(s)J_{\mu}(s)V^{\dag}(s) \label{eq:trans}
\end{eqnarray}
and satisfies the covariant conservation law
\begin{eqnarray*}
\sum_{\mu}D^L_{\mu}J_{\mu}(s)&=& \sum_{\mu}\big(U(s+\mu,\mu)J_{\mu}(s)U^{\dag}(s,\mu)-J_{\mu}(s)\big)\\
&=&0.
\end{eqnarray*}
However it does not satisfy the Abelian conservation law:
\begin{eqnarray}
\sum_{\mu}\big(J_{\mu}(s+\mu)-J_{\mu}(s)\big)= 0. \label{eq:acon}
\end{eqnarray}
Moreover it does not have a property corresponding to the Dirac
quantization condition satisfied by the continuum VNABI, as we assumed.
The last point is very unsatisfactory, since
the topological property as a monopole is essential.

Hence we adopt here the second possibility which can reflect partially the topological property satisfied by VNABI. That is,
we define VNABI on lattice as the Abelian-like monopole\cite{Suzuki:2007jp,Suzuki:2009xy} following
DeGrand and Toussaint\cite{DeGrand:1980eq}. First we define Abelian link and plaquette variables:
\begin{eqnarray}
\theta_{\mu}^a(s)&=&\arctan (U^a_{\mu}(s)/U^0_{\mu}(s))\ \ \ (|\theta_{\mu}^a(s)|<\pi) \label{abel_link}\\
\theta_{\mu\nu}^a(s)&\equiv&\partial_{\mu}\theta_{\nu}^a(s)-\partial_{\nu}\theta_{\mu}^a(s),
\label{abel_proj}
\end{eqnarray}
where $\partial_{\nu}(\partial'_{\nu})$ is a forward (backward) difference.
Then the plaquette variable can be decomposed as follows:
\begin{eqnarray}
\theta_{\mu\nu}^a(s) &=&\bar{\theta}_{\mu\nu}^a(s)+2\pi
n_{\mu\nu}^a(s)\ \ (|\bar{\theta}_{\mu\nu}^a|<\pi),\label{abel+proj}
\end{eqnarray}
where $n_{\mu\nu}^a(s)$ is an integer
corresponding to the number of the Dirac string.
Then VNABI as Abelian monopoles is defined by
\begin{eqnarray}
k_{\mu}^a(s)&=& -(1/2)\epsilon_{\mu\alpha\beta\gamma}\partial_{\alpha}
\bar{\theta}_{\beta\gamma}^a(s+\hat\mu) \nonumber\\
&=&(1/2)\epsilon_{\mu\alpha\beta\gamma}\partial_{\alpha}
n_{\beta\gamma}^a(s+\hat\mu) \nonumber \\
J_{\mu}(s)&\equiv&\frac{1}{2}k_{\mu}^a(s)\sigma^a \label{eq:amon}.
\end{eqnarray}
This definition (\ref{eq:amon}) of VNABI satisfies the Abelian conservation condition (\ref{eq:acon}) and takes an integer value
which corresponds to the magnetic charge obeying the Dirac quantization
condition. The eigenvalue $\lambda_{\mu}$ is defined from (\ref{Eigen}) as 
\begin{eqnarray}
(\lambda_{\mu}(s))^2=\sum_a(k_{\mu}^a(s))^2.\label{Eigen_L}
\end{eqnarray}
However Eq.(\ref{eq:amon}) does not satisfy the transformation property
(\ref{eq:trans}) on the lattice. We will demonstrate that this property is recovered in the continuum limit by showing the gauge invariance of the monopole density or the squared monopole density (\ref{Eigen_L}) in the scaling limit. 

\begin{table}[H]
 \caption{A typical example of monopole loop distributions (Loop length (L) vs Loop number (No.)) for various gauges in one thermalized vacuum on $24^4$
 lattice at $\beta=3.6$ in the tadpole improved action. Here $I$ and $L$ denote the color component and the loop length of the monopole loop, respectively.
 }
 \label{Tab:Mdist}
  \begin{center}
  \begin{tabular}{|c|c||c|c||c|c|}
\hline										
NGF I=1	&		&	MCG I=1	&		&	DLCG I=1	&\\	
\hline										
L	&	No	&	L	&	No	&	L	&	No\\
\hline										
4	&	154	&	4	&	166	&	4	&	164\\
6	&	20	&	6	&	64	&	6	&	66\\
8	&	7	&	8	&	30	&	8	&	28\\
10	&	2	&	10	&	13	&	10	&	15\\
14	&	1	&	12	&	11	&	12	&	10\\
16	&	1	&	14	&	4	&	14	&	3\\
407824	&	1	&	16	&	5	&	16	&	6\\
	&		&	18	&	1	&	18	&	2\\
	&		&	22	&	2	&	20	&	1\\
	&		&	24	&	2	&	22	&	1\\
	&		&	28	&	1	&	24	&	2\\
	&		&	30	&	1	&	26	&	3\\
	&		&	32	&	1	&	30	&	1\\
	&		&	34	&	2	&	36	&	1\\
	&		&	36	&	1	&	44	&	1\\
	&		&	44	&	1	&	48	&	1\\
	&		&	46	&	1	&	54	&	1\\
	&		&	48	&	1	&	58	&	1\\
	&		&	58	&	1	&	124	&	1\\
	&		&	124	&	1	&	1106	&	1\\
	&		&	2254	&	1	&	1448	&	1\\
\hline										
AWL I=1	&		&	MAU1 I=1	&		&	MAU1 I=3	&\\	
\hline										
L	&	No	&	L	&	No	&	L	&	No\\
\hline										
4	&	142	&	4	&	73	&	4	&	190\\
6	&	66	&	6	&	32	&	6	&	80\\
8	&	36	&	8	&	13	&	8	&	22\\
10	&	8	&	10	&	11	&	10	&	15\\
12	&	7	&	12	&	6	&	12	&	2\\
14	&	3	&	14	&	3	&	14	&	3\\
16	&	3	&	16	&	2	&	16	&	1\\
18	&	1	&	18	&	3	&	18	&	3\\
20	&	1	&	20	&	2	&	20	&	3\\
22	&	3	&	22	&	1	&	24	&	1\\
26	&	3	&	30	&	2	&	36	&	1\\
28	&	1	&	34	&	2	&	42	&	1\\
30	&	2	&	58	&	1	&	60	&	1\\
32	&	1	&	148	&	1	&	66	&	1\\
34	&	1	&	5188	&	1	&	146	&	1\\
40	&	1	&		&		&	318	&	1\\
46	&	1	&		&		&	722	&	1\\
58	&	1	&		&		&		&	\\
120	&	1	&		&		&		&\\	
308	&	1	&		&		&		&\\	
1866	&	1	&		&		&		&\\
\hline	
  \end{tabular}
 \end{center}
\end{table}

\subsection{Simulation details}
\subsubsection{Tadpole improved gauge action}
First of all,  we adopt  the tree level improved action of the form \cite{Alford:1995hw} for simplicity in $SU(2)$ gluodynamics:
\beq
    S =   \beta_{imp} \sum_{pl} S_{pl}
       - {\beta_{imp} \over 20 u_0^2} \sum_{rt} S_{rt}
\label{eq:improved_action}
\eeq
where $S_{pl}$ and $S_{rt}$ denote plaquette and $1 \times 2$ rectangular loop terms in the action,
\beq
S_{pl,rt}\ = \ {1\over 2}{\rm Tr}(1-U_{pl,rt}) \, ,
\label{eq:terms}
\eeq
the parameter $u_0$ is the {\it input} tadpole improvement
factor taken here equal to the fourth root of the average plaquette
$P=\langle \frac{1}{2} {\mathrm tr} U_{pl} \rangle$.
In our simulations we have not included one--loop corrections to the coefficients, for the sake of simplicity.

The lattices adopted are  $48^4$ for $\beta=3.0\sim3.9$ and $24^4$  for $\beta=3.3\sim3.9$. The latter was taken mainly for studying finite-size effects.
The simulations with the action (\ref{eq:improved_action})
have been performed with parameters
given in Table~\ref{t1} in Appendix\ref{Ap:tadpole} following similarly the method as
adopted in Ref.\cite{Bornyakov:2005iy}.

\subsubsection{The non-Abelian string tension}
In order to fix the physical lattice scale we need to compute one physical
dimensionful observable the value of which is known. For this purpose we
choose the string tension $\sigma$.
The string tension for the action (\ref{eq:improved_action}) was computed
%% VB I added reference to Bornyakov:2005iy.  I also think it is necessary to add a remark comparing new results and old ones
long ago in \cite{Poulis:1997zx,Bornyakov:2005iy} but we improve this measurement
according to present standards.
We use the hypercubic blocking (HYP) invented by the authors
of Ref.~\cite{Hasenfratz:2001hp,Hasenfratz:2001tw,
Gattringer:2001jf,Bornyakov:2004ii} to reduce the statistical errors.  After one step of HYP, APE smearing \cite{APE} were applied to the space-like links. The spatial
smearing is made, as usually, in order to
variationally improve the overlap with a mesonic flux tube state.
The results of the measured string tensions are listed also in Table~\ref{t1} in Appendix\ref{Ap:tadpole}.
%%%%%%%%%%%%%% Ishiguro

\begin{figure}[htb]
\caption{\label{fig_b-beta} $b=na(\beta)$ in unit of $1/\sqrt{\sigma}$ versus $\beta$}
\includegraphics[width=8cm,height=6.cm]{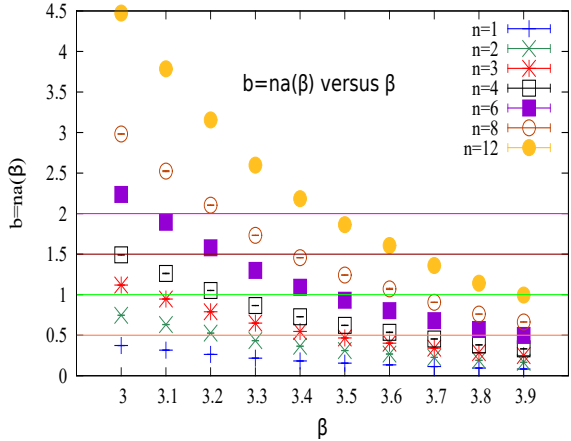}
 \vspace{-0.3cm}
\end{figure}

\begin{table}[h]
 \caption{The $n=4$ blocked monopole loop distribution (Loop length (L) vs Loop number (No.)) in various gauges on $6^4$ reduced
 lattice volume at $\beta=3.6$ in the same vacuum used in Table\ref{Tab:Mdist}.}
 \label{Tab:Mdist3}
   \begin{center}
  \begin{tabular}{|c|c||c|c||c|c|}
\hline											
NGF I=1	&		&	MCG I=1	&		&	DLCG I=1	&		\\
\hline											
L	&	No	&	L	&	No	&	L	&	No	\\
\hline			
9266	&	1	&	4	&	5	&	4	&	8	\\
	&		&	6	&	1	&	6	&	2	\\
	&		&	10	&	1	&	406	&	1	\\
	&		&	340	&	1	&		&		\\
\hline											
AWL I=1	&		&	MAU1 I=1	&		&	MAU1 I=3	&		\\
\hline											
L	&	No	&	L	&	No	&	L	&	No	\\
\hline											
4	&	5	&	4	&	12	&	4	&	8	\\
6	&	1	&	6	&	1	&	6	&	3	\\
14	&	1	&	10	&	1	&	8	&	2	\\
352	&	1	&	24	&	1	&	16	&	1	\\
	&		&	26	&	1	&	276	&	1	\\
	&		&	270	&	1	&		&		\\
\hline											
  \end{tabular}
 \end{center}
\end{table}

\subsubsection{Introduction of smooth gauge-fixings}
Monopole loops in the thermalized vacuum produced in the above improved action (\ref{eq:improved_action}) still contain large amount of lattice artifacts.
Hence we here adopt a gauge-fixing technique smoothing the vacuum, although any gauge-fixing is not necessary in principle in the continuum limit\cite{footnote2}: 
\begin{enumerate}
  \item Maximal center gauge (MCG).\\
The first gauge is the maximal center gauge\cite{DelDebbio:1996mh,DelDebbio:1998uu} which is usually discussed in the framework of the center vortex idea. We adopt the so-called direct maximal center gauge which requires maximization of the quantity
\begin{eqnarray}
R=\sum_{s,\mu}(\tr U(s,\mu))^2 \label{eq:MCG}
\end{eqnarray}
with respect to local gauge transformations.  The condition (\ref{eq:MCG}) fixes the gauge up to
$Z(2)$ gauge transformation and can be considered as the Landau gauge for the adjoint representation. In our simulations, we choose simulated
annealing algorithm as the gauge-fixing method which is known to be  powerful for finding the global maximum. For details, see the reference\cite{Bornyakov:2000ig}.

%Direct 
\item Direct Laplacian center gauge (DLCG).\\
The second is the 
Laplacian center gauge\cite{Faber:2001zs} which is also discussed in connection to center vortex idea. Here we adopt the so-called direct Laplacian center gauge (DLCG).  Firstly, we require maximization of the quantity
\begin{eqnarray}
R_M=\sum_{s,\mu}\tr \left[M^T(s)U^A(s,\mu)M(s,\mu)\right]
\end{eqnarray}
where $U^A(s,\mu)$ denotes the adjoint representation of $U(s, \mu)$ and $M(s,\mu)$ is a real-valued $3\times3$ matrix in $SU(2)$ gauge theory which satisfies the constraint
\begin{eqnarray}
\frac{1}{V}\sum_s\sum_j M^T_{ij}(s)M_{jk}(s)=\delta_{ik}
\end{eqnarray}
with $V$ lattice volume. Matrix field $M(s)$ which leads to a global maximum of $R_M$ is composed of the three lowest eigenfunctions of a lattice Laplacian operator.  Secondly, to determine the corresponding gauge transformation, we construct $SO(3)$ matrix-valued field which is the closest to $M(s)$ and satisfies the corresponding Laplacian condition by local gauge transformation. Finally,  the $SO(3)$ matrix-valued field is mapped to an SU(2) matrix-valued field which is used to the gauge transformation for the original lattice gauge field in fundamental representation. After that, 
%LCG-->DLCG
DLCG maximizes the quantity (\ref{eq:MCG}) with respect to solving a lattice Laplacian equation.

\item Maximal Abelian Wilson loop gauge (AWL).\\
Another example of a smooth gauge is introduced. It is the maximal Abelian Wilson loop gauge (AWL) in which
\begin{eqnarray}
R&=&\sum_{s,\mu\neq\nu}\sum_a(cos(\theta^a_{\mu\nu}(s)) \label{SAWL}
\end{eqnarray}
is maximaized. Here
$\theta^a_{\mu\nu}(s)$  have been introduced in eq.~(\ref{abel+proj}).
Since $cos(\theta^a_{\mu\nu}(s))$ are $1\times 1$ Abelian Wilson loops, the gauge is called as the maximal Abelian Wilson
loop gauge (AWL). A similar gauge was proposed in \cite{Suzuki:1996ax}, although only one-color component was considered then
in comparison with the maximal Abelian gauge (MAG). Note that even $1\times 1$ small Abelian Wilson loop is enhanced when a smooth gauge
condition such as the MA gauge is adopted. The details are presented in the Appendix \ref{AWL}.

\item Maximal Abelian and $U(1)$ Landau gauge (MAU1).\\
The fourth is the combination of the maximal Abelian gauge (MAG) and the $U(1)$ Landau gauge\cite{Kronfeld:1987ri,Kronfeld:1987vd}. Namely we first perform the maximal Abelian gauge fixing and then with respect to the remaining $U(1)$ symmetry the Landau gauge fixing is done. This case breaks the global $SU(2)$ color symmetry contrary to the previous three cases (MCG, 
%LCG-->DLCG
DLCG and AWL) but nevertheless we consider this case since the vacuum is smoothed fairly well. MAG is the gauge which maximizes
\begin{equation}
  R=\sum_{s,\hat\mu}{\rm Tr}\Big(\sigma_3 U(s,\mu)
        \sigma_3 U^{\dagger}(s,\mu)\Big) \label{R}
\end{equation}
with respect to local gauge transformations. Then there remains $U(1)$ symmetry to which the Landau gauge fixing is applied, i.e.,
$ \sum_{s,\mu} cos \theta^3_{\mu}(s)$   is maximized\cite{Bali:1996dm}.
\end{enumerate}

\vspace{0.5cm}

\subsubsection{Extraction of infrared monopole loops}
An additional improvement is obtained when we extract important long monopole clusters only from total monopole loop distribution. Let us see a
 typical example of monopole loop distributions in each gauge in comparison with that without any gauge fixing starting from a thermalized vacuum at $\beta=3.6$ on $24^4$ lattice. They are shown in Table \ref{Tab:Mdist}.
One can find almost all monopole loops are connected and total loop lengths are very large when no gauge fixing (NGF)
%(NGF) inserted
 is applied as shown
in the NGF case.
On the other hand, monopole loop lengths become much shorter in all smooth gauges discussed here.
Also it is found that only one or few loops are long enough and others are very short as observed similarly in old papers in MAG.
The long monopole clusters are called as infrared monopoles and they are
the key ingredient giving confinement as shown in the old papers\cite{Ejiri:1994uw}.
It is important that in addition to MAU1, all other three MCG, DLCG and AWL cases also have similar behaviors. Since small separate monopole loops can be regarded as
lattice artifacts, we extract only infrared monopoles alone. Although there observed only one infrared monopole loop in almost all cases, there are some vacua (especially for large beta) having  two or three separate long loops which can be seen as infrared one, since they have much longer length than other shorter ones. We here define as infrared monopoles as all loops having loop lengths longer than $10\%$ of the longest one. The cutoff value is not so critical. Actually the definition of infrared loops itself has an ambiguity, since even in the longest loop, we can not separate out some short artifact loops attached accidentally to the real infrared long loop. But such an ambiguity gives us numerically only small effects as seen from the studies of different cutoff values.
\begin{figure}[htb]
\caption{The VNABI (Abelian-like monopoles) density versus $a(\beta)$ in MCG on $48^4$. Top: total density; bottom: infrared density. $n^3$ in the legend means
$n$-step blocked monopoles.}
  \begin{minipage}[b]{0.9\linewidth}
    \centering
    \includegraphics[width=8cm,height=6.cm]{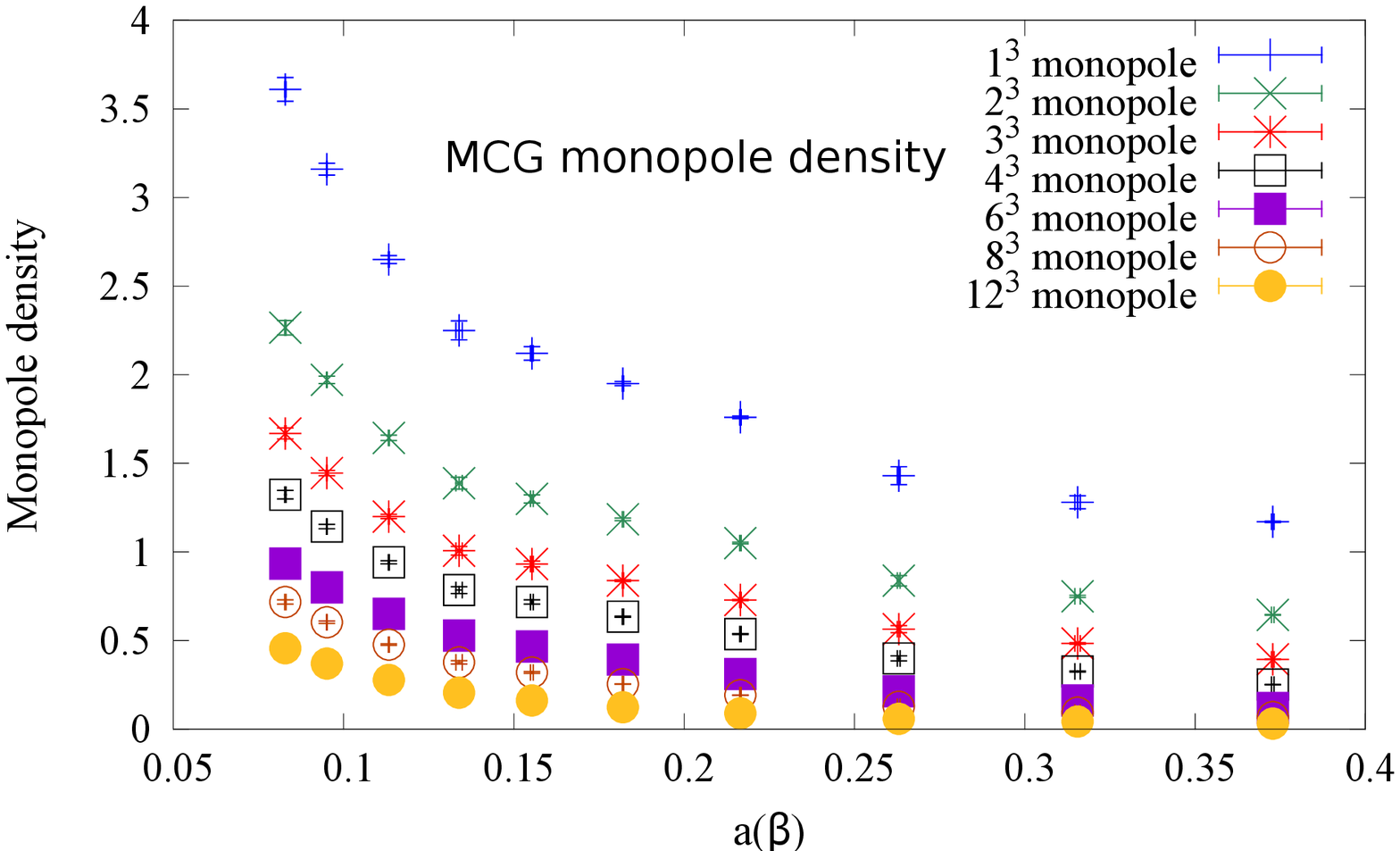}
    \label{fig_MCG-a}
  \end{minipage}
  \begin{minipage}[b]{0.9\linewidth}
    \centering
    \includegraphics[width=8cm,height=6.cm]{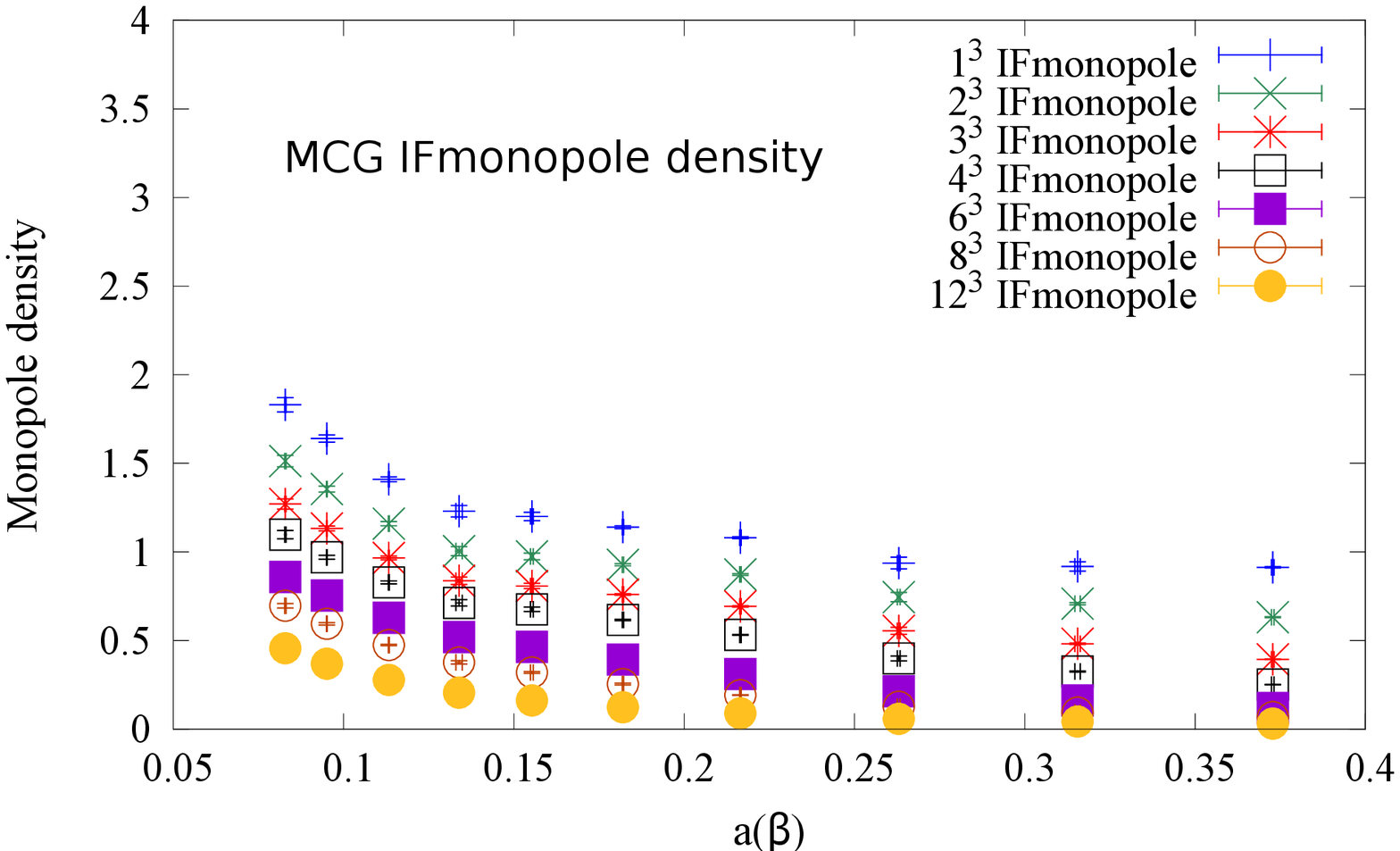}
    \label{fig_IF_MCG-a}
  \end{minipage}
\end{figure}

\begin{figure*}[tbh]
\caption{The VNABI (Abelian-like monopoles) density versus $b=na(\beta)$ in MCG on $48^4$. Top: total density; bottom: infrared density.}
  \begin{minipage}[b]{0.9\linewidth}
    \centering
    \includegraphics[width=14cm,height=10.cm]{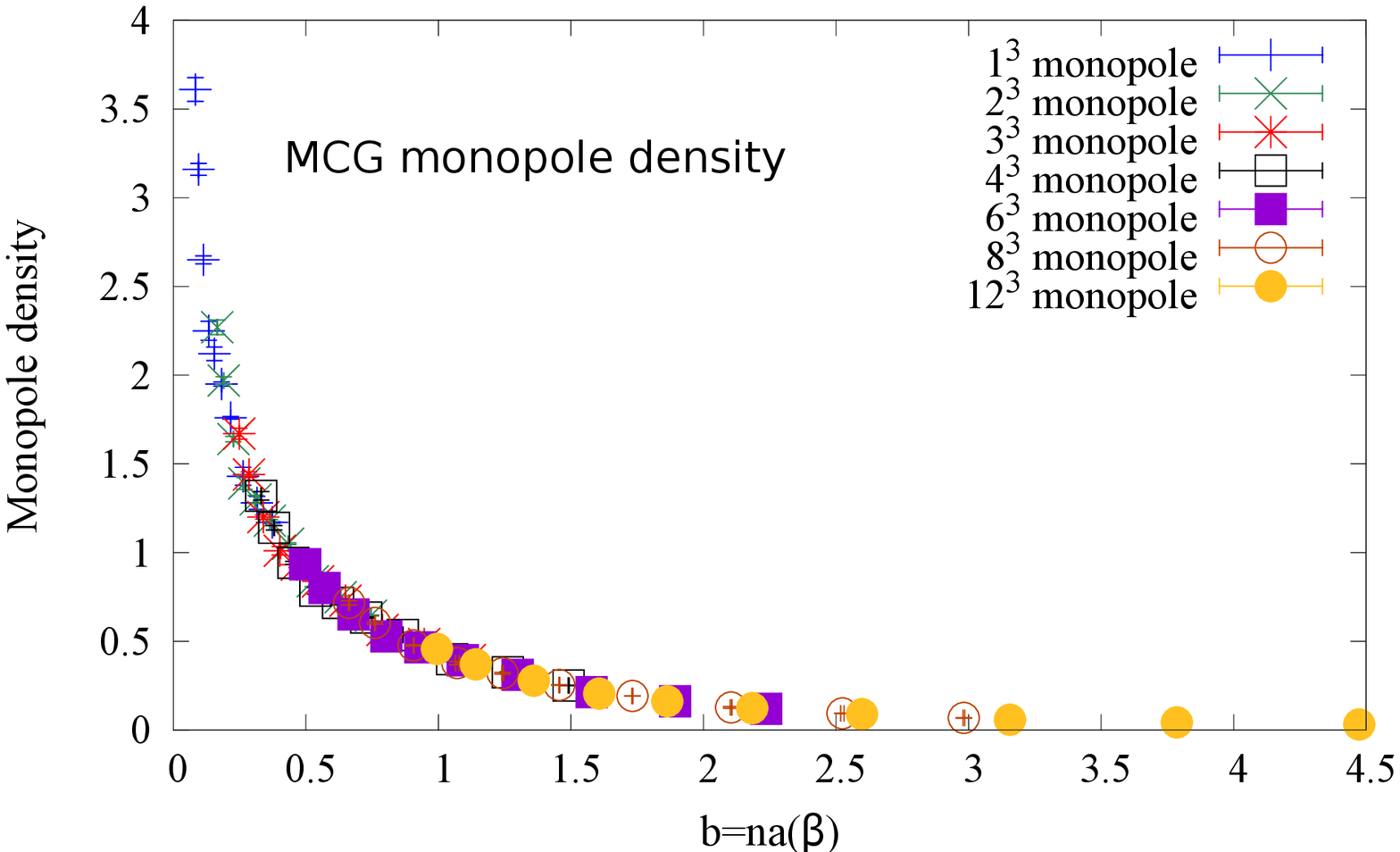}
    \label{fig_MCG-b}
  \end{minipage}
  \begin{minipage}[b]{0.9\linewidth}
    \centering
 \includegraphics[width=14cm,height=10.cm]{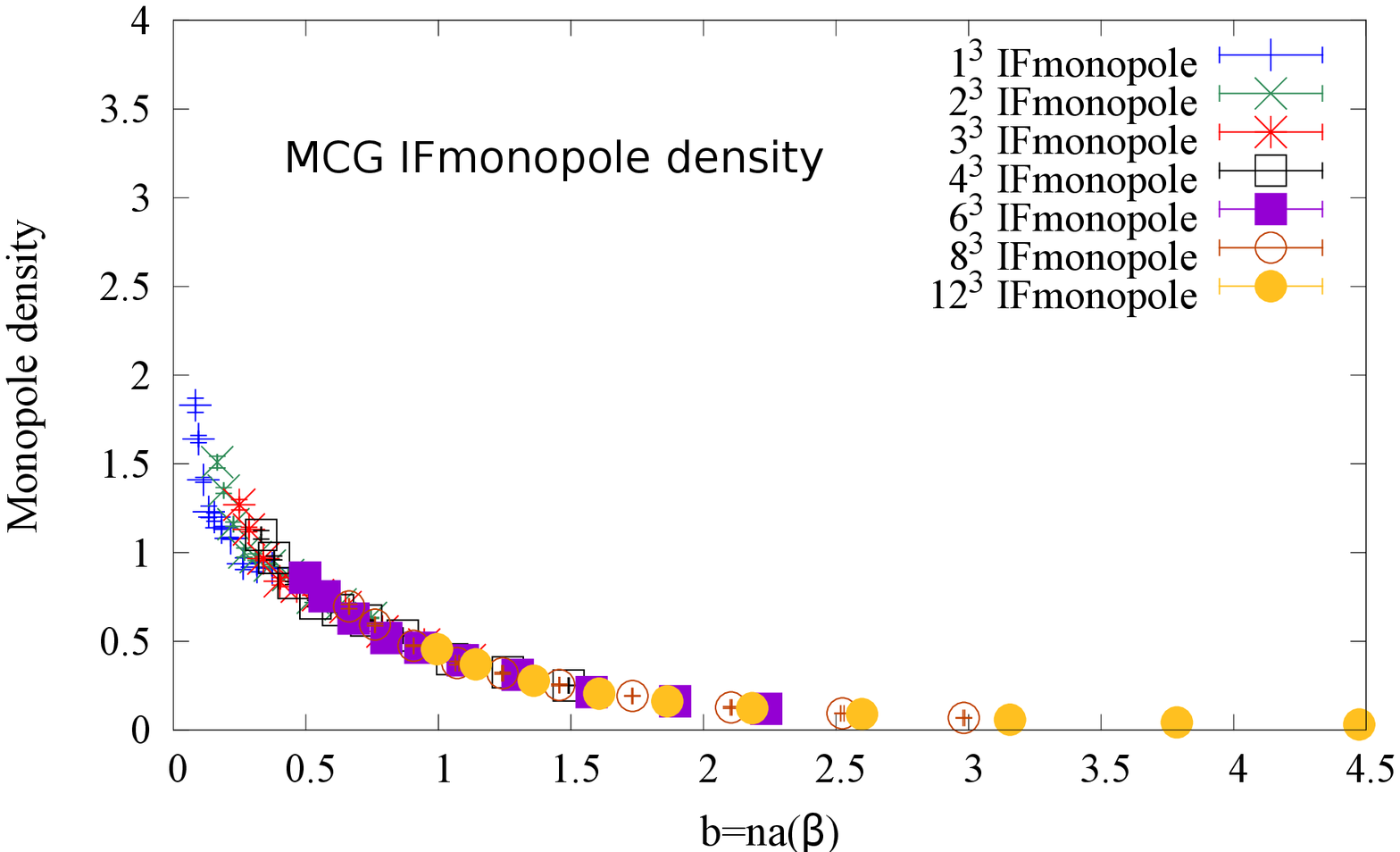}
    \label{fig_IF_MCG-b}
  \end{minipage}
\end{figure*}

\begin{figure*}[hbt]
\caption{The fit of the infrared VNABI (Abelian-like monopoles) density data in MCG on $48^4$ lattice to Eq.(\ref{eq:rho-b}). }
  \begin{minipage}[b]{0.9\linewidth}
    \centering
    \includegraphics[width=10cm,height=7.cm]{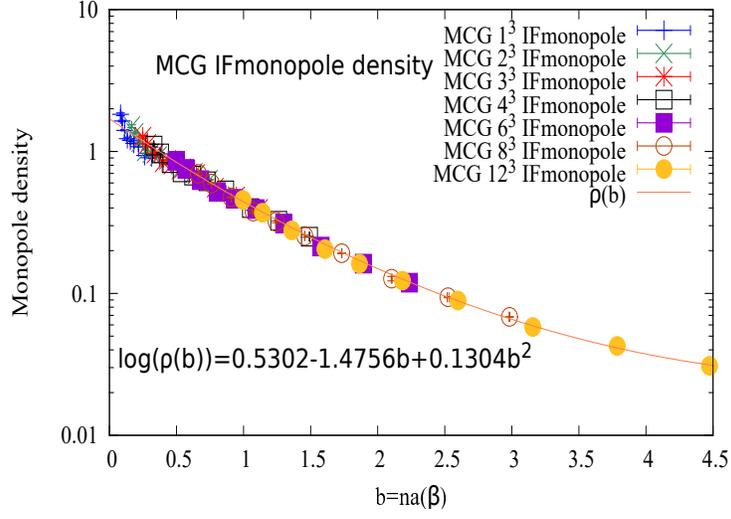}
    \label{fig_f(b)}
  \end{minipage}
\end{figure*}

\begin{figure*}[hbt]
\caption{The VNABI (Abelian-like monopole) density at $b=0.5, 1.0, 1.5, 2.0$  for different $n$ in MCG on $48^4$. The data used are derived by a linear interpolation of two nearest data below and above  for the corresponding $b$  and $n$. As an example, see the original data at $b=1.0$ in Table\ref{Tab_b=1}.}
  \begin{minipage}[b]{0.9\linewidth}
    \centering
    \includegraphics[width=12cm,height=7.cm]{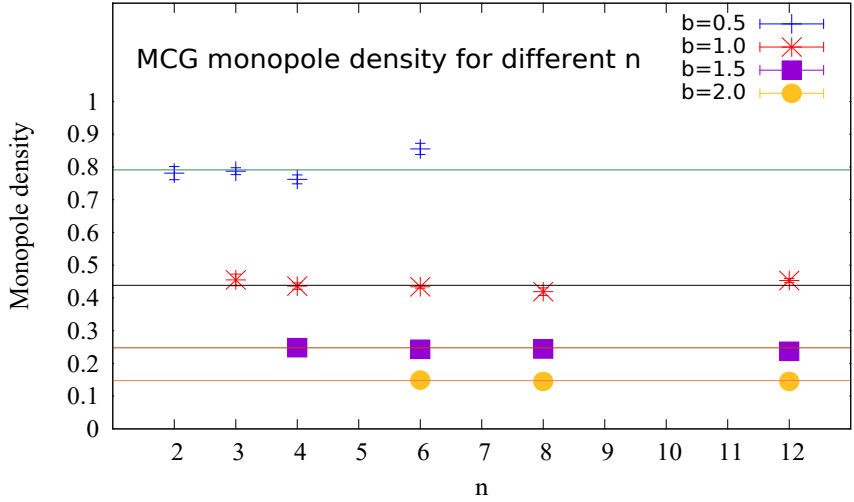}
    \label{fig_MCG-b1}
  \end{minipage}
\end{figure*}

\begin{table}[htbp]
 \caption{IF monopole density $\rho_{IF}$ around $b=1.0$ for each blocking steps $n$ in MCG case on $48^4$.\label{Tab_b=1}}% 
 \begin{center}
  \begin{tabular}{|r|r|r|r|r|r|} 
    \hline
    $n$   & $\beta$   & $b=na(\beta)$   & $db$   &  $\rho_{IF}$  & error   \\
    \hline
3	& 3.0	& 1.1184	& 0.0012	& 3.94E-01	& 1.42E-03 \\
3	& 3.1	& 0.9465	& 0.0024	& 4.82E-01	& 4.06E-03 \\
    \hline
4	& 3.2	& 1.052	& 0.0016	& 3.99E-01	& 1.40E-02 \\
4 	& 3.3	& 0.866	& 0.0008	& 5.32E-01	& 2.37E-03 \\
    \hline
6	& 3.4	& 1.092	& 0.0012	& 3.93E-01	& 2.80E-03 \\
6 	& 3.5	& 0.9318	& 0.0024	& 4.64E-01	& 7.44E-03 \\
    \hline
8	& 3.6	& 1.0712	& 0.0072	& 3.77E-01	& 9.20E-03 \\
8 	& 3.7	& 0.9064	& 0.0008	& 4.75E-01	& 3.78E-03 \\
    \hline
12	& 3.8	& 1.1412	& 0.0012	& 3.70E-01	& 4.43E-03 \\
12 	& 3.9	& 0.9948	& 0.0024	& 4.56E-01	& 8.36E-03 \\
    \hline
  \end{tabular}
 \end{center}
\end{table}

\begin{figure}[htb]
\caption{The VNABI (Abelian-like monopoles) density versus $b=na(\beta)$ in AWL on $48^4$. Top: total density; bottom: infrared density.}
  \begin{minipage}[b]{0.9\linewidth}
    \centering
    \includegraphics[width=8cm,height=6.cm]{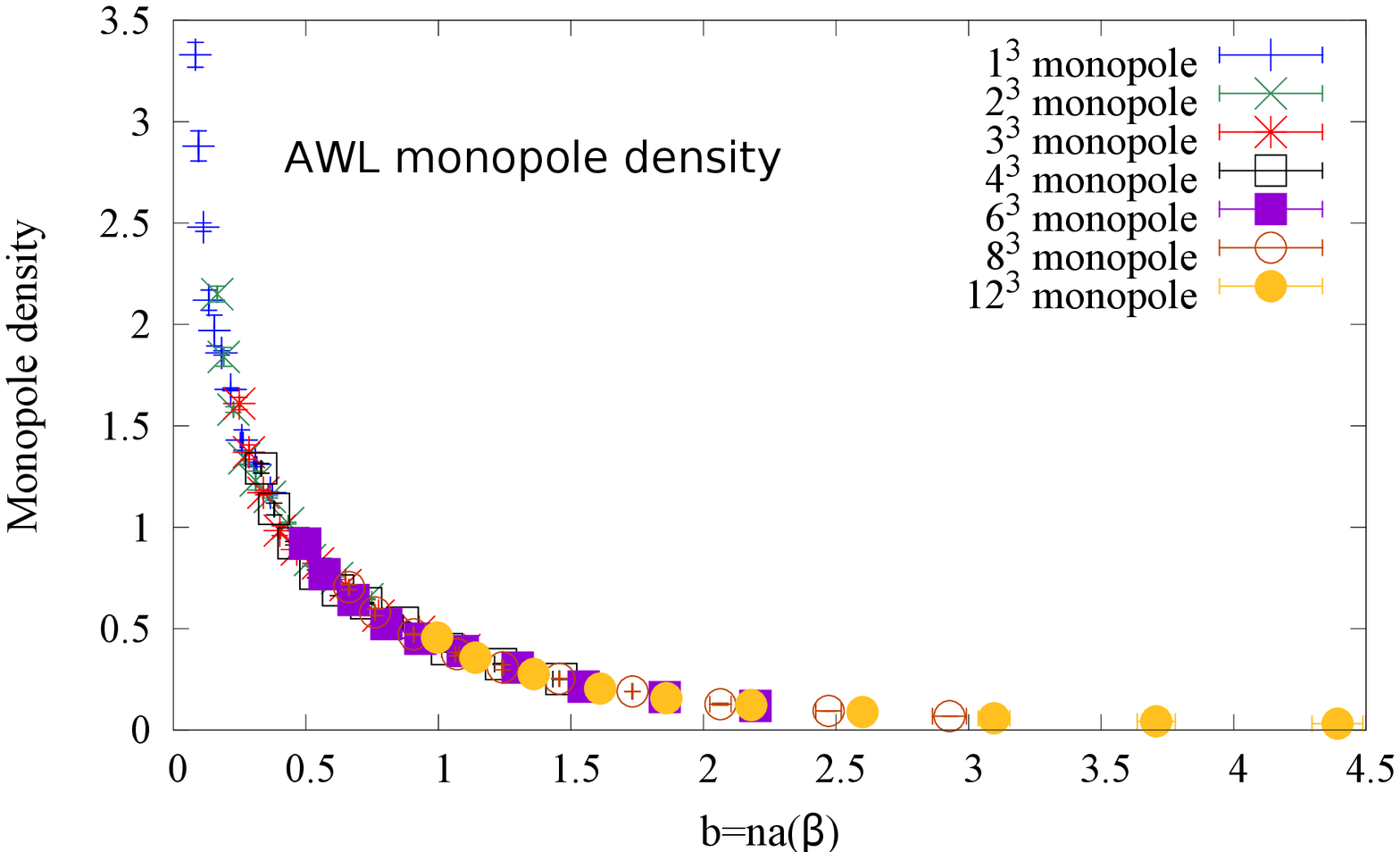}
    \label{fig_AWL-b}
  \end{minipage}
  \begin{minipage}[b]{0.9\linewidth}
    \centering
    \includegraphics[width=8cm,height=6.cm]{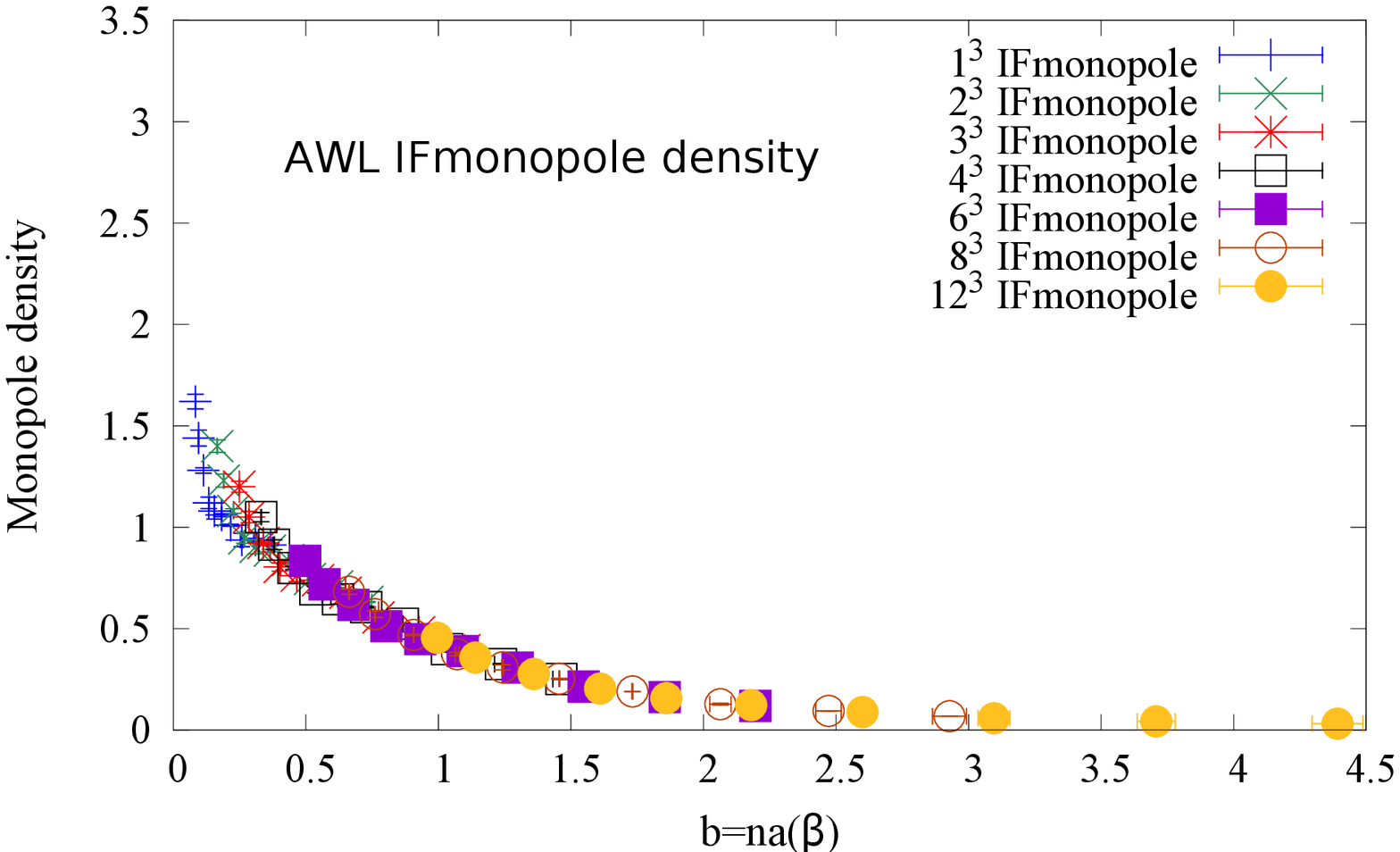}
    \label{fig_IF_AWL-b}
  \end{minipage}
\end{figure}

\begin{figure}[hbt]
\caption{The VNABI (Abelian-like monopoles) density versus $b=na(\beta)$ in DLCG on $24^4$.  }
  \begin{minipage}[b]{0.9\linewidth}
    \centering
    \includegraphics[width=8cm,height=6.cm]{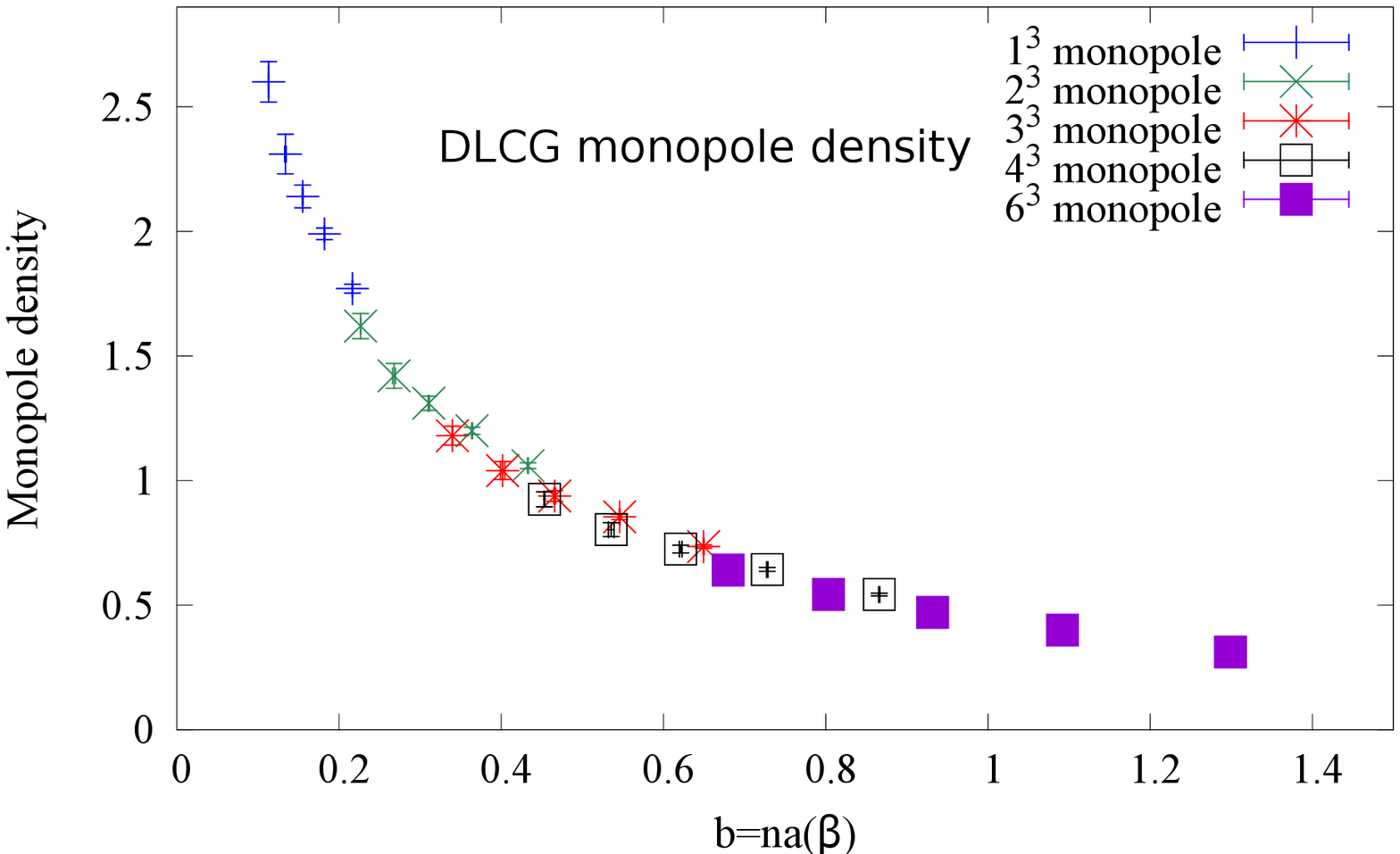}
    \label{fig_DLCG-b}
  \end{minipage}
\end{figure}

\subsubsection{Blockspin transformation}
Block-spin transformation and the renormalization-group method is known as the powerful tool to study the continuum limit. We introduce the blockspin transformation with respect to Abelian-like monopoles. The idea was first introduced by Ivanenko et al.\cite{Ivanenko:1991wt} and applied in obtaining an infrared effective monopole action in Ref.\cite{Shiba:1994db}. The $n$ blocked monopole has a total magnetic charge inside the $n^3$ cube and is defined on
a blocked reduced lattice with the spacing $b=na$,  $a$ being the spacing of the original lattice. The respective magnetic currents are defined as
\begin{eqnarray}
k_{\mu}^{(n)}(s_n) &=& \frac{1}{2}\epsilon_{\mu\nu\rho\sigma}
\partial_{\nu}n_{\rho\sigma}^{(n)}(s_n+\hat{\mu}) \nonumber\\
%% VB question: what is 'ns' here?
%% TS answer: here s is the site number on the reduced lattice and
%%  ns+... is the site number on the original lattice.
    & = & \sum_{i,j,l=0}^{n-1}k_{\mu}(ns_n \nonumber\\
    &&  +(n-1)\hat{\mu}+i\hat{\nu}
     +j\hat{\rho}+l\hat{\sigma}), \label{excur}\\
n_{\rho\sigma}^{(n)}(s_n) &=& \sum_{i,j=0}^{n-1}
n_{\rho\sigma}(ns_n+i\hat{\rho}+j\hat{\sigma}),\nonumber
\end{eqnarray}
where $s_n$ is a site number on the reduced lattice.
For example,
\begin{eqnarray*}
 k_{\mu}^{(2)}(s_2)&=&
\sum_{i,j,l=0}^{1}k_{\mu}(2s_2+\hat{\mu}+i\hat{\nu}
     +j\hat{\rho}+l\hat{\sigma}),\\
k_{\mu}^{(4)}(s_4)&=&\sum_{i,j,l=0}^{3}k_{\mu}(4s_4+3\hat{\mu}+i\hat{\nu}
     +j\hat{\rho}+l\hat{\sigma}) \nonumber \\
            &=&\sum_{i,j,l=0}^{1}k_{\mu}^{(2)}(2s_4+\hat{\mu}+i\hat{\nu}
     +j\hat{\rho}+l\hat{\sigma}).
\end{eqnarray*}
These equations show that the relation between $k_{\mu}^{(4)}(s_4)$ and $k_{\mu}^{(2)}(s_2)$ is similar to that between $k_{\mu}^{(2)}(s_2)$ and $k_{\mu}(s)$ and hence one can see the above equation (\ref{excur}) corresponds to the usual block-spin transformation.
After the block-spin transformation, the number of short lattice artifact  loops decreases while loops having larger magnetic charges appear. We show an example of the loop length and loop number distribution of the four step ($n=4$ )
blocked monopoles in Table\ref{Tab:Mdist3} with respect to the same original vacuum as in Table\ref{Tab:Mdist}. For reference, we show the relation between
the spacing of the blocked lattice and $\beta$ in Fig.\ref{fig_b-beta}.
In Fig.1 and in what follows we present  spacings $a$ and $b$ in units of $1/\sqrt{\sigma}$.

\begin{figure}[hbt]
\caption{The VNABI (Abelian-like monopoles) density versus $b=na(\beta)$  for $k^2$ and $k^3$ components in MAU1 on $48^4$. Top: total density; bottom: infrared density.\label{fig_MA_k23}}
  \begin{minipage}[b]{0.9\linewidth}
    \centering
    \hspace*{-1cm}
    \includegraphics[width=9cm,height=6.cm]{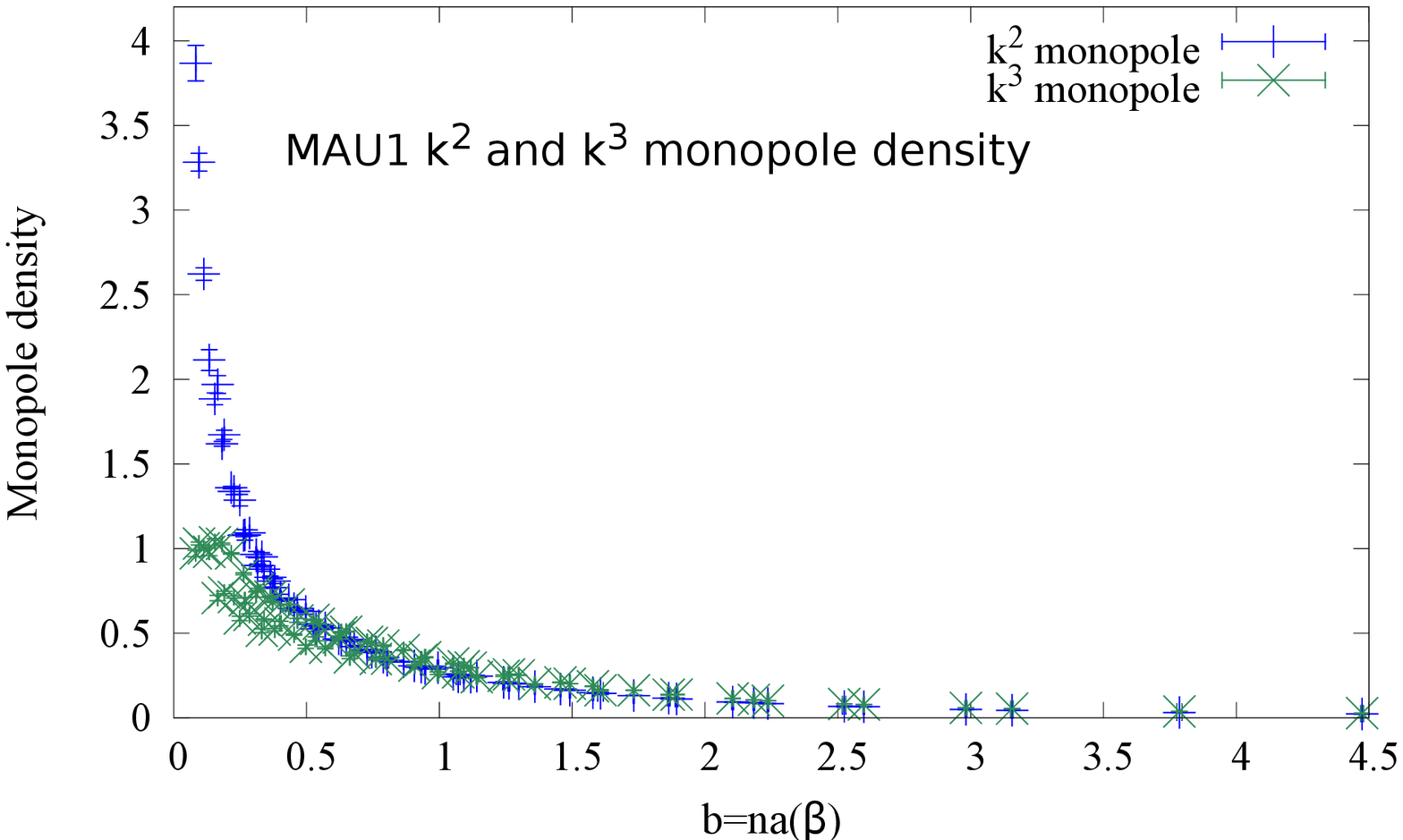}
      \end{minipage}
  \begin{minipage}[b]{0.9\linewidth}
    \centering
    \hspace*{-1cm}
    \includegraphics[width=9cm,height=6.cm]{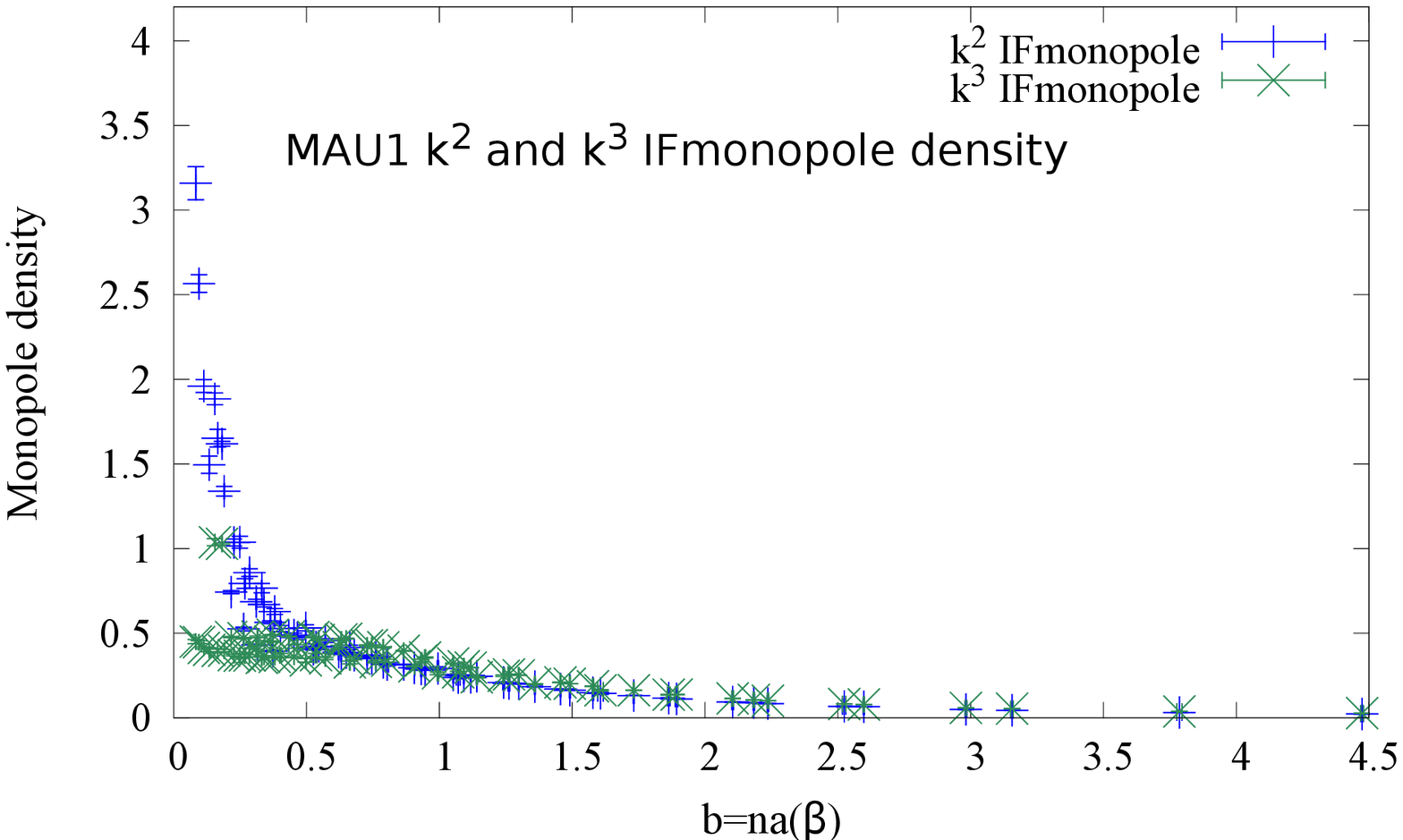}
  \end{minipage}
\end{figure}

\subsection{Numerical results}
Now let us show the simulation results with respect to VNABI (Abelian-like monopole ) densities. Since monopoles are three-dimensional objects, the density is defined as follows:
\begin{eqnarray}
\rho=\frac{\sum_{\mu,s_n}\sqrt{\sum_a(k_{\mu}^a(s_n))^2}}{4\sqrt{3}V_nb^3},\label{eq:Mdensity}
\end{eqnarray}
where $V_n=V/n^4$ is the 4 dimensional volume of the reduced lattice, $b=na(\beta)$ is the spacing of the reduced lattice after $n$-step blockspin transformation.
$s_n$ is the site on the reduced lattice and the superscript $a$ denotes a color component. Note that $\sum_a(k_{\mu}^a)^2$
is gauge-invariant in the continuum limit. Although the global color invariance is exact except in MAU1 gauge, the average of the density of each
color component of $|k_{\mu}^a|$ is not equal to the average of the above $\rho$, since two or three colored monopoles can run on
the same dual links. In general, the density $\rho$ is a function of two variables $\beta$ and $n$.

\subsubsection{Scaling}
For the purpose of studying the continuum limit, it is usual to analyse scaling behaviors.
 First of all, let us show the data of MCG case in Fig.\ref{fig_MCG-a}. In this Figure and in what follows we present the monopole density $\rho$ in units of $\sigma^{1.5}$. When the scaling exists for both the string tension and the monopole density, we expect $\rho\to\textrm{const}$ as $a(\beta)\to 0$ and $V\to\infty$, since $a(\beta)$ is measured in unit of the string tension.
In the case of total monopole density
such a behavior is not seen yet. When infrared monopoles alone and blocked monopoles are considered, the behavior becomes flatter
as seen from Fig.\ref{fig_IF_MCG-a}. But still this scaling is not  conclusive.
We need to study larger $\beta$ regions on larger lattice volumes. These features are  very much
similar in other smooth gauges
as 
%MAW
AWL, DLCG and MAU1 and so their data are not shown here.

\subsubsection{Scaling under the block-spin transformations}
It is very interesting to see that
more beautiful and clear scaling behaviors are observed when we plot $\rho(a(\beta),n)$ versus $b=na(\beta)$. As one can see from the figures shown below
for various smooth gauges considered in this work, one
can see a universal function $\rho(b)$ for $\beta=3.0\sim3.9$ ($\beta=3.3\sim3.7$) and $n=1,2,3,4,6,8,12$ ($n=1,2,3,4,6$) on $48^4$ ($24^4$) lattice. Namely \textit{$\rho(a(\beta),n)$ is a function of $b=na(\beta)$ alone.} 
Thus we observe clear indication of the continuum ($a(\beta)\to 0$) limit for the lattice VNABI studied in this work.

\begin{figure*}[htb]
\caption{The VNABI (Abelian-like monopoles) density (\ref{eq:Mdensity}) versus $b=na(\beta)$  in MAU1 on $48^4$. Top: total density; bottom: infrared density.\label{fig_MA-b}}
  \begin{minipage}[b]{0.9\linewidth}
    \centering
    \includegraphics[width=12cm,height=8.cm]{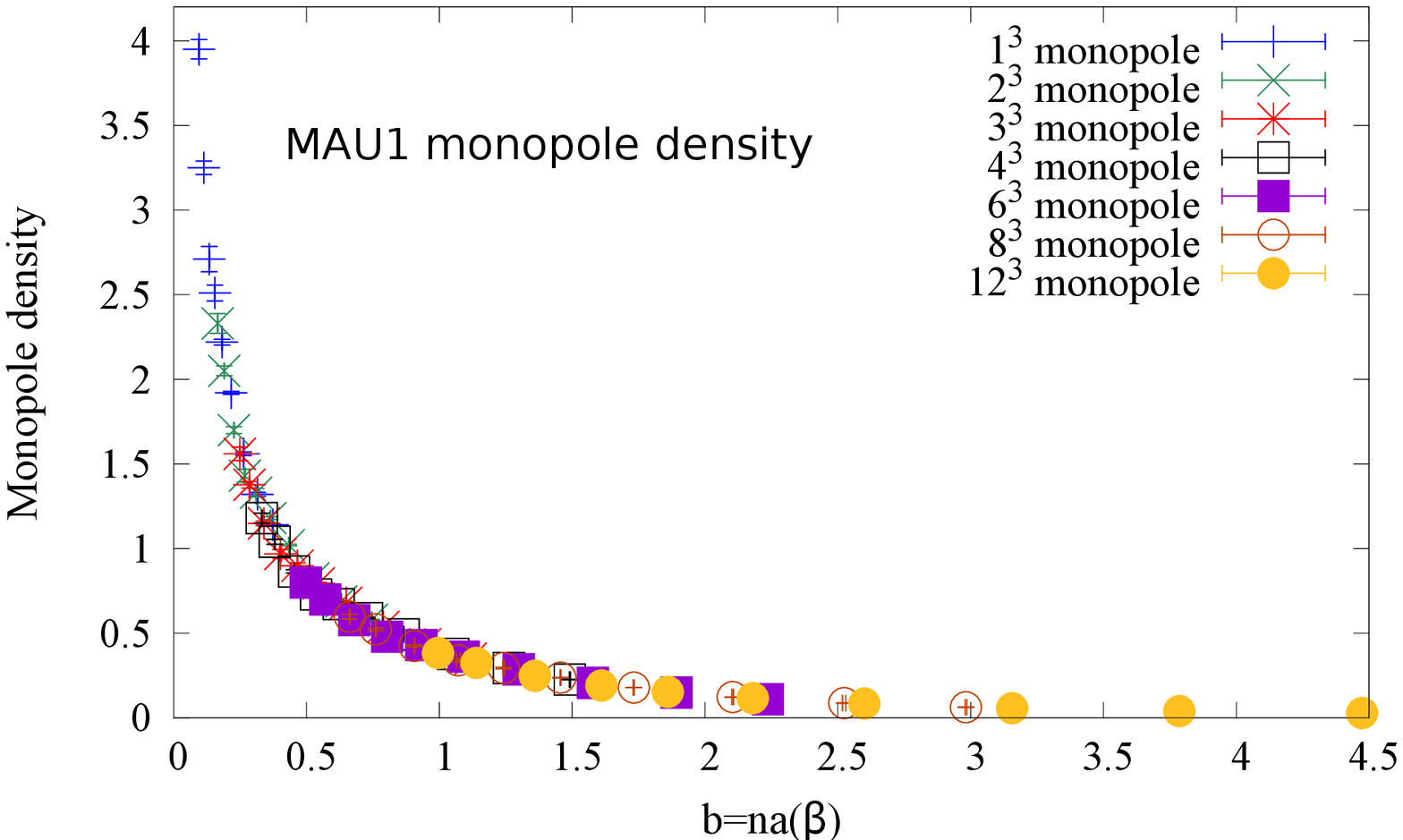}
  \end{minipage}
  \begin{minipage}[b]{0.9\linewidth}
    \centering
    \includegraphics[width=12cm,height=8.cm]{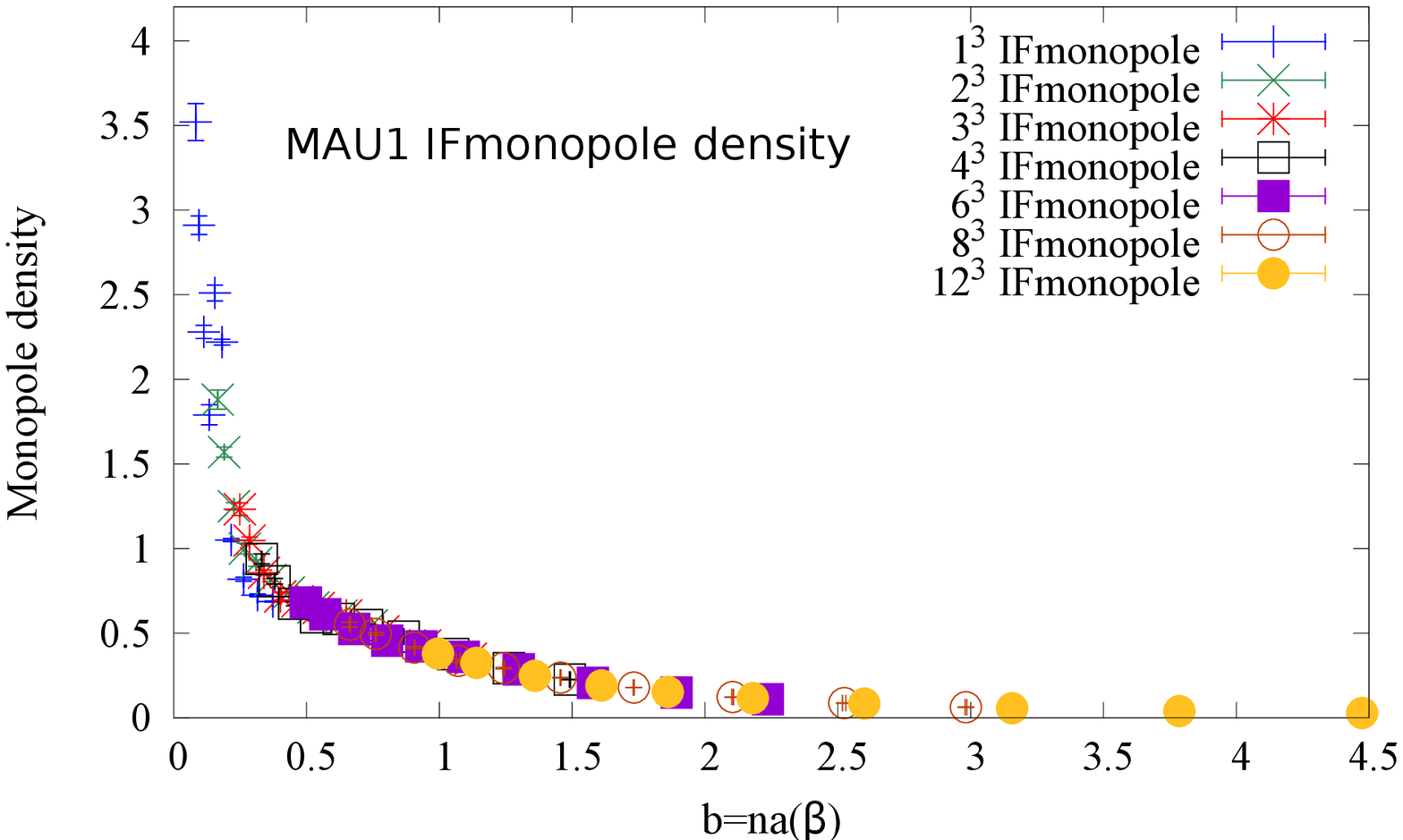}
      \end{minipage}
\end{figure*}

\subsubsection{MCG case}
First we show the case of MCG gauge-fixed vacua in details. As can be seen from
Fig.\ref{fig_MCG-b}, data for $\rho(a(\beta),n)$ can be expressed by a function of  one argument $b=na(\beta)$ alone. There is a very beautiful scaling behavior for the range of $\beta=3.0\sim3.9$  and $n=1,2,3,4,6,8,12$. When we are restricted to long infrared monopoles alone, the density becomes substantially reduced for small $b<0.5$ region. But the scaling also can be seen except for small $b$ region as shown in Fig.\ref{fig_IF_MCG-b}. The violation of scaling for small $b$ region is mainly due to the ambiguity of extracting infrared monopoles. When we restrict ourselves to the data for $b\ge 0.5$, the scaling function $\rho(b)$ is obtained using the $\chi^2$ fit to a simple function as shown in Fig.\ref{fig_f(b)}: 
\begin{eqnarray}
\rho(b)&=&\exp(a_1 + a_2b + a_3b^2),\label{eq:rho-b}\\
a_1&=& 0.5302(141), a_2=-1.4756(158), a_3= 0.1304(35). \nonumber
\end{eqnarray}
But the fit  is not good enough, since $\chi^2/N_{dof}=12.56$ for $N_{dof}=44$. Here we show the function (\ref{eq:rho-b}) only for the purpose of illustration, since we have not found a simple but better fit. 

To see  in more details, let us consider the data points at $b=0.5, 1.0, 1.5, 2.0$ for each $n$. Especially the data at $b=1.0$ can be fixed from the data at 5 different values of $\beta$ from $3.0\le\beta\le 3.9$  as seen from Fig.\ref{fig_b-beta} and Table\ref{Tab_b=1}.
One can see the scaling behavior also clearly from the density plot for 
different $n$ at  $b=1.0, 1.5, 2.0$ as shown in Fig.\ref{fig_MCG-b1}. However  a scaling violation is seen at $b=0.5$\cite{footnote3}.  
%correction footnote2-->footnote3

\subsubsection{AWL case}
Very similar behaviors are seen in the AWL gauge case. Again beautiful scaling behaviors for the range of $\beta=3.0\sim3.9$  and $n=1,2,3,4,6,8,12$ are seen in Fig.\ref{fig_AWL-b}. But in the case of infrared monopoles shown in Fig.\ref{fig_IF_AWL-b}, a scaling violation is observed for small $b$ region.

\subsubsection{DLCG case}
Since the DLCG gauge-fixing needs much time for larger lattice, we evaluate monopole density only on $24^4$ lattice.
As seen from Fig.\ref{fig_DLCG-b}, a scaling behavior is found, although small deviations exist for small $b$ region.
%% VB remark
% there is data set in DLCG  (presumably at beta=3.5) which is not on the universal curve. I think this should be checked
%% VB end
\subsubsection{MAU1 case}
Now we discuss the case of MAU1 gauge.
In this gauge, the global isospin symmetry is broken.
Hence let us first evaluate the monopole density in each color direction. Namely
\begin{eqnarray}
\rho^a=\frac{\sum_{\mu,s_n}|k_{\mu}^a(s_n))|}{4V_nb^3}.
\end{eqnarray}
As expected we find $\rho^1\sim\rho^2\neq\rho^3$, so that we show $\rho^2$ and $\rho^3$.
The results are shown in Fig.\ref{fig_MA_k23}.  Here the scaling is seen clearly with respect to the off-diagonal $k^2$ currents, but the violation is seen for the diagonal $k^3$ currents especially at small $b$ region.
Similar behaviors are found when we are restricted to infrared monopoles.

However when we evaluate the monopole density (\ref{eq:Mdensity}), we can observe  similar beautiful scaling behaviors as in MCG and AWL cases. They are shown in
Fig.\ref{fig_MA-b}. 

%\vspace{.5cm}
\begin{figure*}[htb]
\caption{Comparison of the VNABI (Abelian-like monopoles) densities versus $b=na(\beta)$ in MCG, AWL, DLCG and MAU1 cases. DLCG data only are on $24^4$ lattice. Here $\rho(b)$ is a scaling function (\ref{eq:rho-b}) determined from the Chi-Square fit to the IF monopole density data in MCG. Top: total density; bottom: infrared density. \label{fig_log-b}}
  \begin{minipage}[b]{0.9\linewidth}
    \centering
    \includegraphics[width=12cm,height=8.cm]{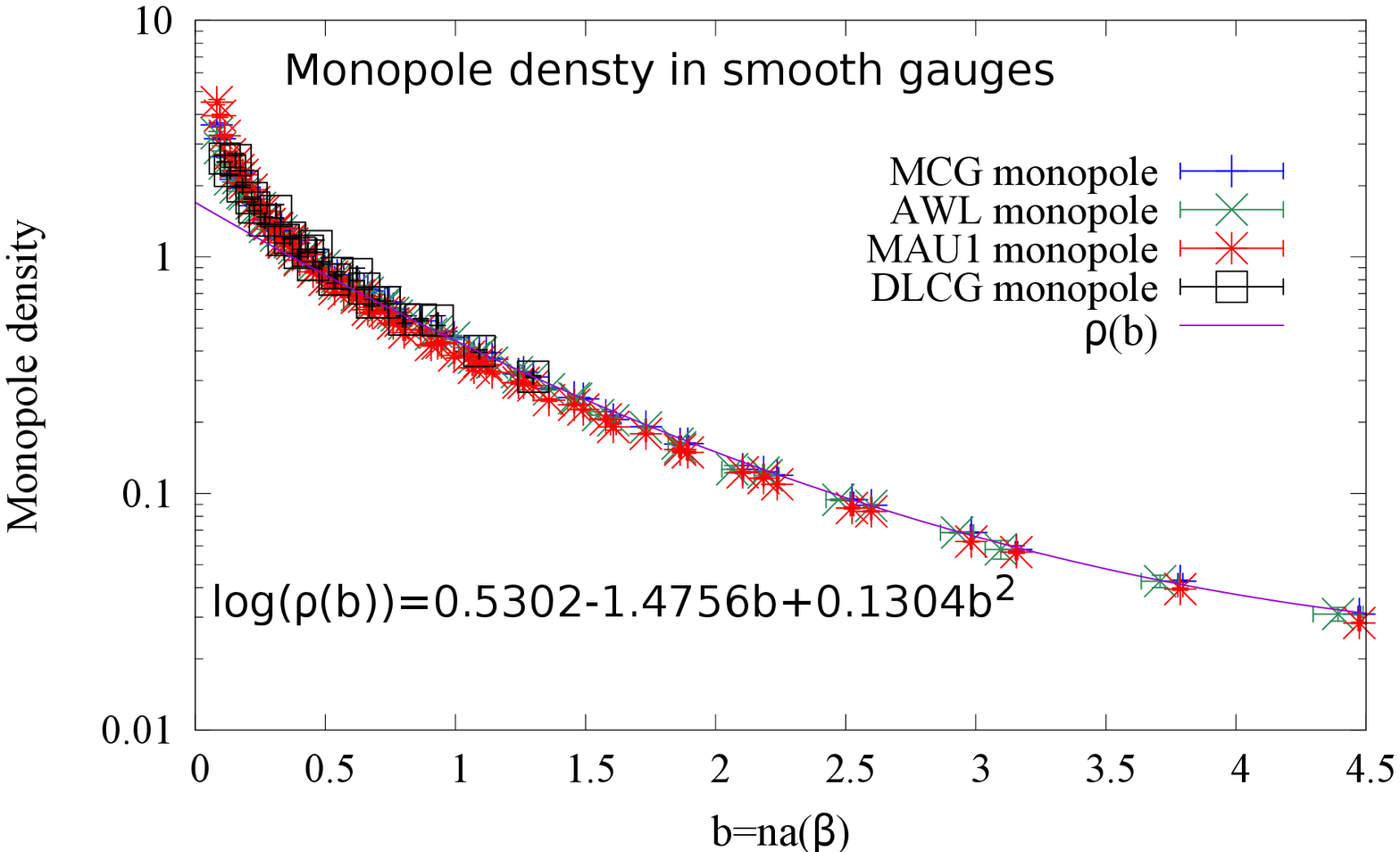}
     \end{minipage}
  \begin{minipage}[b]{0.9\linewidth}
    \centering
    \includegraphics[width=12cm,height=8.cm]{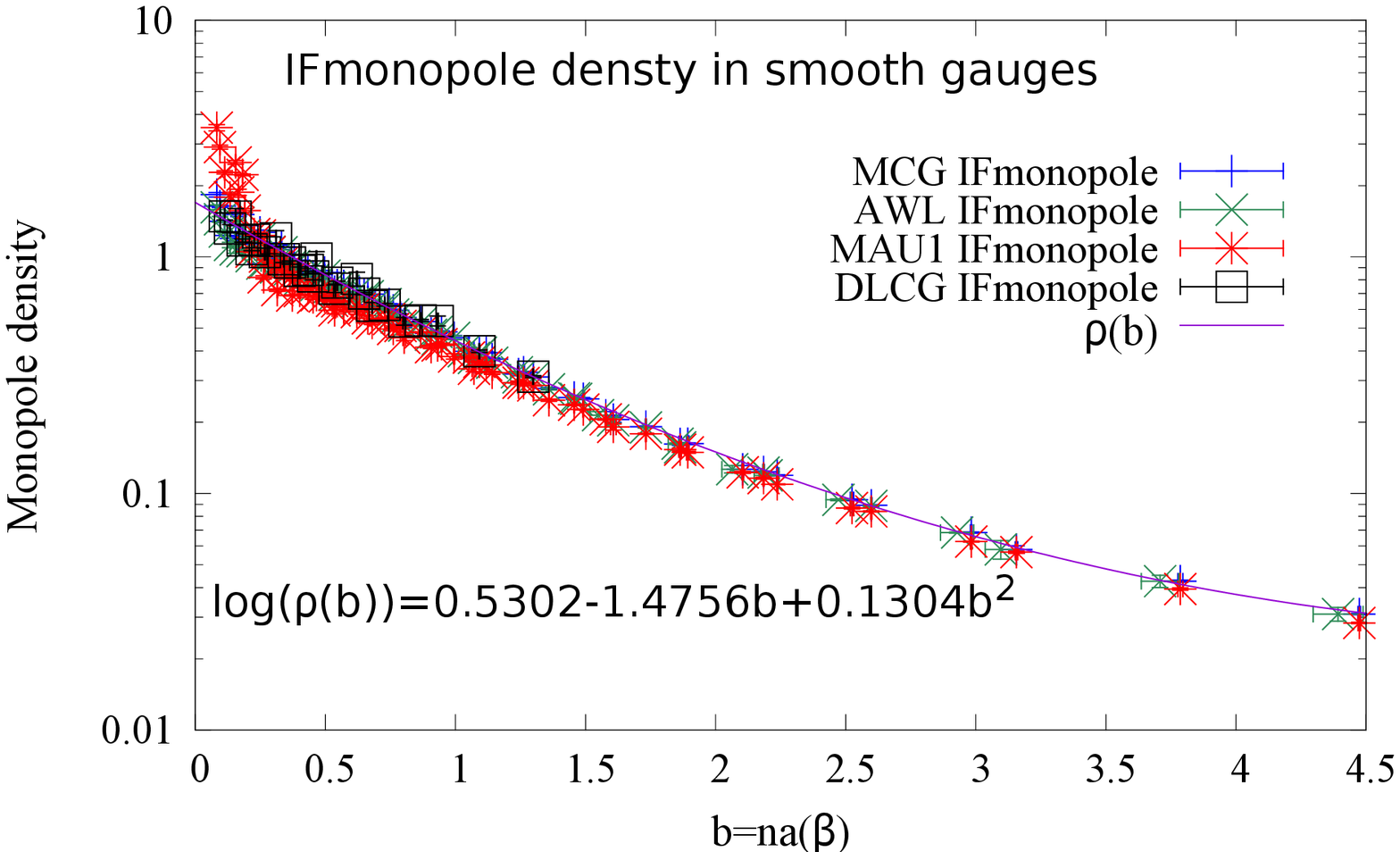}
  \end{minipage}
\end{figure*}

\begin{figure}[htb]
\caption{\label{fig_MCG3} Volume dependence of VNABI (Abelian-like monopole) density in
the case of MCG in $48^4$ and $24^4$ tadpole improved gauge action.  The data for $3.0\le\beta\le 3.6$ and $1\le n\le 6$ alone are plotted for comparison.}
\includegraphics[width=8cm,height=6.cm]{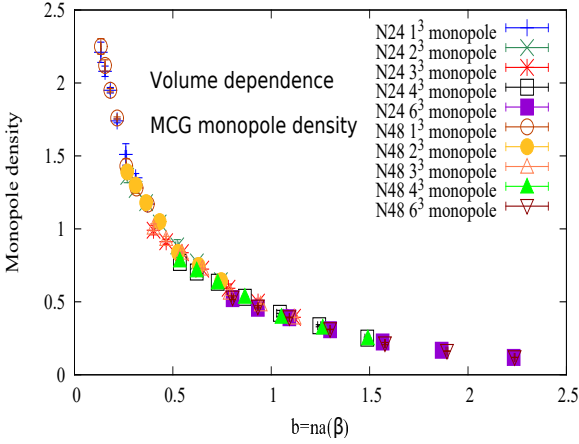}
 \vspace{-0.3cm}
\end{figure}

\begin{figure}[htb]
\caption{\label{fig_impWil} Gauge action dependence of VNABI (Abelian-like monopole) densities in
the case of DLCG in $24^4$ tadpole improved and Wilson gauge actions,  The data for $3.3\le\beta\le 3.7$ and $1\le n\le 6$ alone are plotted.}
\includegraphics[width=8cm,height=6.cm]{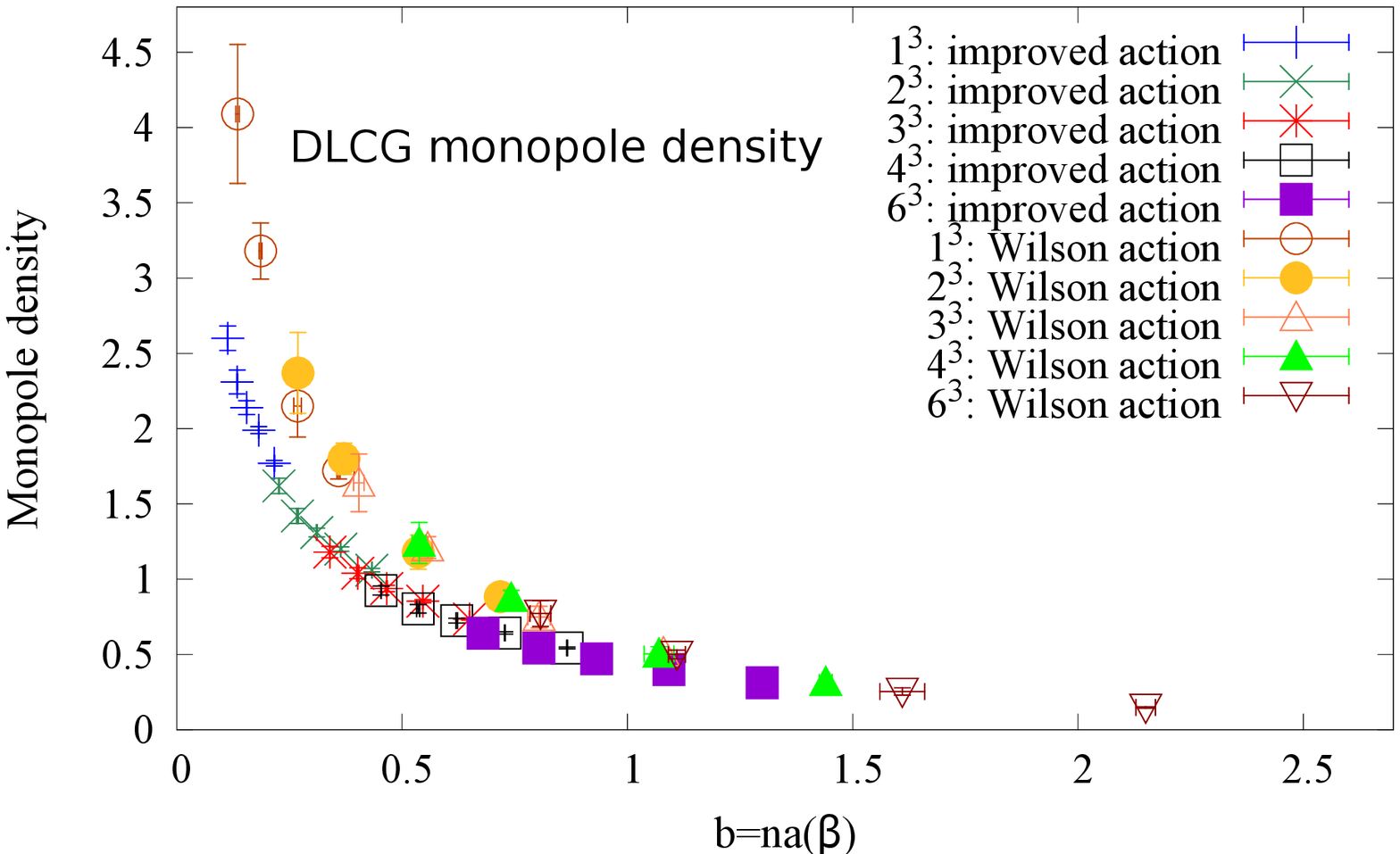}
 \vspace{-0.3cm}
\end{figure}

\subsection{Gauge dependence}
Since $\sum_a(k_{\mu}^a)^2$ should be gauge-invariant according to our derivation in section \ref{Sec2}, we compare the data in different smooth gauges. Look at Fig.{\ref{fig_log-b}}, which show the comparison of the data in  four gauges (MCG, AWL, DLCG and MAU1).  One can see that data obtained in these four different gauges are in good agreement with each other providing strong indication of gauge independence. \textit{This is the main result of this work.}  Note that in MAU1 gauge, the global color invariance is broken and usually off-diagonal color components of gauge fields are said to have large lattice artifacts. However here we performed additional U1 Landau gauge-fixing with respect to the remaining $U(1)$ symmetry after MA  fixing, which seems to make the vacua smooth enough as those in MCG gauge case. The fact that the scaling functions $\rho(b)$ obtained in MCG gauge can reproduce other three smooth-gauge data seems to show that it is near to the smallest density corresponding to the continuum limit without large lattice artifact effects. In other non-smooth gauges 
%such as NGF
or without any gauge-fixing (NGF),  $\rho$ does not satisfy the scaling and actually  becomes much larger. This is due to our inability to suppress lattice artifacts in the non-smooth gauges or without gauge-fixing.

\subsection{Volume dependence in MCG case}
The volume dependence is also studied when the two data on $48^4$ and $24^4$ lattices in MCG are plotted for the same $\beta$ region $(3.0\le\beta\le3.6)$ and the blocking steps $(1\le n\le 6)$ as shown in Fig.{\ref{fig_MCG3}}. We found sizable finite volume effects for $\beta=3.7$ only (not shown in the figure) when lattice size for $L=24$ becomes $La  < 2.7/ \sqrt{\sigma} $. Volume dependence for 
$(3.0\le\beta\le3.6)$ is very small as seen from Fig.{\ref{fig_MCG3}}.

\subsection{Gauge action dependence\label{sec8}}
Let us in short check how the gauge action adopted here improves the density $\rho$ behavior by comparing the data in the tadpole improved action with those in the simple Wilson gauge action. It is shown in Fig.\ref{fig_impWil}.   The density in the Wilson action is higher especially for $b\le 1.0$ and so considerable improvement is obtained with the choice of the tadpole improved gauge action.

\section{Conclusions}
In conclusion, we have proposed a new color confinement scheme which is summarized as follows:
\begin{enumerate}
  \item VNABI is equal to the Abelian-like monopole coming from the violation of the Abelian-like Bianchi identities.
  \item VNABI satisfies the Abelian-like conservation law as well as the covariant one. Hence there are $N^2-1$ conserved magnetic charges in the case of color $SU(N)$.
%%% deleted 
\item   All magnetic charges  are assumed to satisfy the Dirac quantization condition.
%quantized \`{a} la Dirac.
  \item VNABI can be defined on lattice as lattice Abelian-like monopoles. Previous numerical results suggest that the dual Meissner effect due to condensation of VNABI must be the color confinement mechanism of QCD. The role of Abelian monopoles is played by VNABI. This must be a new scheme for color confinement in QCD. 
%%added
  \item VNABI are assumed to satisfy $[J_\mu, J_{\nu\neq\mu}]=0$ leading to the simultaneous diagonalization for all $\mu$.
%%%
  \item Condensation of the color invariant  magnetic currents $\lambda_{\mu}$ which are the eigenvalue of VNABI $J_{\mu}$ 
may be a key mechanism of  the physical confining vacuum.
\end{enumerate}

Then to check if the new confinement scenario is correct in the continuum limit, densities of VNABI defined on lattice were studied extensively in this work. Since VNABI is equivalent to Abelian-like monopoles in the continuum, VNABI on lattice is defined as lattice Abelian-like monopoles following DeGrand-Toussaint\cite{DeGrand:1980eq}. This definition even on lattice keeps  partially the topological property of VNABI satisfied in the continuum.

In the thermalized vacuum, there are plenty of lattice artifact monopoles which contribute equally to the density, so that
we have adopted various improvement techniques reducing the lattice artifacts.
One of them is to adopt the tadpole improved gauge action.  The second is to introduce various gauges smoothing the vacuum, although gauge-fixing is not necessary at all in the continuum. We have considered here four smooth gauges, MCG, DLCG, AWL and MAU1.  The third is to perform a blockspin renormalization group study.

With these improvement techniques, we have been able to get very beautiful results.
First of all, in MCG, AWL and MAU1 gauges, clear scaling behaviors are observed up to the 12-step blockspin transformations for $\beta=3.0\sim 3.9$. Namely the density $\rho(a(\beta),n)$ is a function of $b=na(\beta)$ alone, i.e. $\rho(b)$.
If such scaling behaviors are seen for $n\to\infty$, the obtained curve depending on $b=na(\beta)$ alone corresponds
to the continuum limit $a(\beta)\to 0$.
It is just the renormalized trajectory.
The second beautiful result is the gauge independence of the measured densities at least with respect to MCG, AWL and MAU1 smooth gauges on $48^4$ and DLCG on $24^4$ adopted here. The gauge independence is the property expected in the continuum limit, since the observed quantity $\rho$ in (\ref{eq:Mdensity}) is gauge invariant in the continuum.

These beautiful results suggest that the lattice VNABI adopted here has the continuum limit and hence the new confinement scenario can
 be studied on lattice with the use of the lattice VNABI.

   \vspace{.5cm}

Let us note that monopole dominance and the dual Meissner effect due to VNABI as Abelian monopoles were   shown partially without any smooth gauge fixing with the use of random gauge transformations in Ref.\cite{Suzuki:2007jp,Suzuki:2009xy}, although  scaling behaviors were not studied enough. More extensive studies of these effects and derivation of infrared effective VNABI action using block-spin transformation in these smooth gauges discussed here and its application to analytical studies of non-perturbative quantities will appear in near future.

\begin{acknowledgments}
The numerical simulations of this work were done using  computer clusters HPC and SX-ACE at Reserach Center for Nuclear Physics (RCNP) of Osaka University
and the supercomputer at ITEP, Moscow.
%%% and Moscow computer%%%
The authors would like to thank RCNP for their support of computer facilities. Work of VB was supported by  Russian Foundation for Basic Research (RFBR) grant 16-02-01146.
One of the authors (T.S.) would like to thank Prof. T. Kugo  and Prof. H. Tamura for pointing him the errors in the original paper and fruitful discussions.
\end{acknowledgments}

\appendix
\section{Tadpole improved action\label{Ap:tadpole}}
The parameter $u_0$
has been iterated over a series of Monte Carlo runs in order to match the
fourth root of the average plaquette $P$. The values of $u_0$ are shown in Table \ref{t1}.

\begin{table}[htb]
\begin{center}
\caption{Details of the simulations with improved action}\label{t1}
\vspace{.3cm}
\setlength{\tabcolsep}{0.55pc}
\begin{tabular}{cccccc}
$\beta_{imp}$ & L & $N_{conf}$ & $u_0$ & $<P>^{1/4} $ & $\sqrt{\sigma a^2}$ \\
\hline
3.0 & 24 &  100 & 0.89485 & 0.89510(3) & 0.372(3)  \\
3.0 & 48 &  50 & 0.89485 & 0.89478(1) & 0.3728(4)  \\
3.1 & 24 &  100 &0.90069  & 0.90097(4) & 0.311(2)  \\
3.1 & 48 &  50 & 0.90069 & 0.900688(1) & 0.3155(8)  \\
3.2 & 24 &  100 &0.90578  & 0.90601(3) &  0.261(4) \\
3.2 & 48 &  50 & 0.90578 & 0.905762(1) &  0.2630(4) \\
3.3 & 24 & 100 & 0.910151 & 0.910152(2) & 0.220(2)  \\
3.3 & 48 &  50 & 0.910151 & 0.910150(1) & 0.2165(2)  \\
3.4 & 24 & 100 & 0.91402 & 0.914021(1) & 0.1822(5)  \\
3.4 & 48 &  50 & 0.91402 & 0.914017(1) & 0.1822(1)  \\
3.5 & 24 & 100 & 0.917475 & 0.917480(1) & 0.1555(6)  \\
3.5 & 48 &  50 & 0.917475 & 0.917478(1)& 0.1546(3)  \\
3.6 & 24 & 100 & 0.920616 & 0.920616(1) &0.1306(3)\\
3.6 & 48 &  50 & 0.920616 & 0.920615(1) & 0.1308(1)  \\
3.7 & 24 & 100 & 0.92349 & 0.917484(2) &  0.1124(3)\\
3.7 & 48 &  50 & 0.92349 & 0.923484(1) & 0.1122(1)  \\
3.8 & 48 &  50 & 0.926120 & 0.926126(1) & 0.0951(1)  \\
3.9 & 48 &  50 & 0.928548 & 0.928573(1) & 0.0829(2)  \\
\hline
\end{tabular}
\end{center}
\end{table}

\section{The maximal Abelian Wilson loop gauge\label{AWL}}

In the maximal Abelian Wilson loop gauge (AWL),
\begin{eqnarray}
R&=&\sum_{s,\mu\neq\nu}\sum_a(cos(\theta^a_{\mu\nu}(s)) \label{SAWL}
\end{eqnarray}
is maximized. Here $\theta_{\mu\nu}^a(s)$ is defined in Eq.(\ref{abel_proj}).

Since the gauge transformation property of the Abelian link fields is not simple, to do the gauge-fixing efficiently is not easy. Hence we adopt a gauge fixing iteration method of a minimal gauge transformation starting from the already-known smooth gauge configurations such as those in the maximal center gauge (MCG) or the direct Laplacian center gauge (DLCG) where the quantity $R$ in (\ref{SAWL}) is known to be already large.

At the site $s$, the minimal gauge transformation is written as
\begin{eqnarray*}
U'(s,\mu)&=&e^{i\vec{\alpha}(s)\cdot\vec{\sigma}}U(s,\mu)\\
&=&(1+i\vec{\alpha}(s)\cdot\vec{\sigma})U(s,\mu)+O((\vec{\alpha})^2).
\end{eqnarray*}
Hence in case of the minimal gauge transformation, we get
\begin{eqnarray*}
U'_0(s,\mu)&=&U_0(s,\mu)-\vec{\alpha}(s)\cdot\vec{U}(s,\mu)\\
\vec{U'}(s,\mu)&=&\vec{U}(s,\mu)+U_0(s,\mu)\vec{\alpha}(s)-\vec{\alpha}\times \vec{U}(s,\mu).
\end{eqnarray*}

Then an Abelian link field (\ref{abel_link}) is transformed as
\begin{eqnarray*}
\theta^{'a}_{\mu}(s)&=&\theta^{a}_{\mu}(s)+\delta_{\mu}^a(s),\\
&&\\
\delta_{\mu}^a(s)&=&\alpha^a(s)\\
&+&\frac{1}{(U^0(s,\mu))^2+(U^a(s,\mu))^2}\\
&\times&\bigl(U^a(s,\mu)\sum_{b\neq a}\alpha^b(s)U^b(s,\mu)\\
&&-\epsilon_{abc}U^0(s,\mu)U^c(s,\mu)\bigr).
\end{eqnarray*}

The function $R$ is changed as follows:
\begin{eqnarray*}
R'&=&\sum_{a,\mu\neq\nu,s}cos(\theta_{\mu\nu}^{a'}(s))\\
&=&\sum_{a,\mu\neq\nu,s}cos(\theta_{\mu\nu}^{a}(s)+\delta_{\mu}^a(s)
-\delta_{\nu}^a(s))\\
&=&R-\sum_{a,\mu\neq\nu,s}(\delta_{\mu}^a(s)
-\delta_{\nu}^a(s))sin(\theta_{\mu\nu}^{a}(s))\\
&=&R-\sum_b\alpha^b(s)A^b(s)
\end{eqnarray*}

\newpage

\begin{eqnarray*}
A^b(s)&=&2\sum_{a\neq b}\sum_{\mu\neq\nu}(U^b(s,\mu)
-\epsilon_{bca}U^c(s,\mu))\\
&\times&\frac{U^0(s,\mu)sin(\theta_{\mu\nu}^{a}(s))}{U^0(s,\mu))^2+(U^a(s,\mu))^2}.
\end{eqnarray*}
Hence if we choose
%\vspace{-1cm}
\begin{eqnarray*}
\alpha^b(s)&=&-cA^b(s)\ \ \ (c>0),
\end{eqnarray*}
%\vspace{-1cm}
we get
%\vspace{-1cm}

\begin{eqnarray*}
R'=R+c\sum_b(A^b(s))^2\ge R.
\end{eqnarray*}
%\vspace{-1cm}

The maximum value of $R$ is $3.0$. Actually $R$ in MCG gauge for $\beta=3.3$ is around $2.508$. When the parameter $c$ is taken as small as $0.005$, $R$ becomes $R\sim 2.512$ after four iterations and then tends to decrease. It is the vacuum adopted as the AWL vacuum. If we start from the thermalized vacuum without any smooth gauge-fixing, the large value of $R$ is not obtained with this minimal gauge transformation method.

%\vspace*{3cm}


\begin{thebibliography}{10}
\bibitem{footnote1}
Applying  the Coleman-Mandula theorem  to QCD, we note that Kugo-Ojima\cite{Kugo} derived a manifestly covariant and local canonical operator formalism of non-Abelian gauge theories. Although introducing an indefinite metric is inevitable, the unitarity of the physical $S$-matrix is proved.  Moreover the string-like behavior existing in  gauge fields in the case of color confinement is not rejected in the framework.
\bibitem{footnote2}
The reader may wonder why  a cooling or a smearing method smoothing the vacuum is not used 
instead of introducing a gauge-fixing. 
But these methods do not keep the value of the gauge action, changing 
the real vacuum. Hence they are not considered here. %
\bibitem{footnote3}
%
The reduced chisquare $\chi^2/N_{dof}$ of the fit at $b=0.5$ is $6.26071$ for $N_{dof}=3$, whereas
those at $b=1.0,\ 1.5\ \textrm{and}\ 2.0$ are $2.52175\ (N_{dof}=4)$, $\ 2.80242\ (N_{dof}=3)$ and 
$2.25911\ (N_{dof}=2)$, respectively.
\end{thebibliography}
\end{document}